\documentclass[aip,jcp,reprint,showpacs,longbibliography,noeprint]{revtex4-1} 
\usepackage[utf8]{inputenc}

\usepackage{cmap} 
\usepackage[T1]{fontenc} 

\usepackage{amsmath,amssymb,mathtools,amsbsy}
\IfFileExists{mtpro2.sty}
{ 
  \usepackage{times}
  \usepackage[lite,eucal,subscriptcorrection,slantedGreek,zswash]{mtpro2}
}{
\usepackage{txfonts}

}

\usepackage{mleftright}
\usepackage{microtype}
\usepackage{graphicx}
\usepackage{grffile,subfigure}
\newlength{\figwidth}
\usepackage{url,hyperref}
\usepackage[table,usenames,dvipsnames]{xcolor}
\hypersetup{colorlinks=true, linkcolor=BrickRed,
  urlcolor=blue!50!black, citecolor=blue!50!black}

\usepackage[capitalize]{cleveref}
\crefrangelabelformat{equation}{(#3#1#4)$-$(#5#2#6)}
\let\ref\cref

\usepackage{textcomp}

\usepackage{tikz} %

\newcommand{\kb}{k_{\rm B}}
\renewcommand{\vec}[1]{\mathbf{#1}}

\newcommand{\upd}{\mathrm{d}}

\newcommand{\DHBM}{D_{\rm HBM}}

\definecolor{myblue}{RGB}{94.3418,129.735, 181.708}
\definecolor{myokker}{RGB}{225.465, 156.426, 36.3651}
\definecolor{mygreen}{RGB}{143.406, 177.042, 49.8906}
\definecolor{myorange}{RGB}{236.167, 98.7203, 53.5498}
\definecolor{mypurple}{RGB}{135.293, 120.48, 179.546}
\definecolor{mybrown}{RGB}{197.652, 110.478, 26.2111}
\definecolor{mycyan}{RGB}{93.1579, 158.336, 200.281}
\definecolor{myyellow}{RGB}{255, 192, 0}
\definecolor{mymagenta}{RGB}{165.792, 96.809, 157.193}

\usepackage[normalem]{ulem}

\begin{document}

\title{Orientational dynamics of a heated Janus particle}

\author{Dipanjan Chakraborty}
\email{chakraborty@iisermohali.ac.in}
\affiliation{
  Indian Institute of Science Education and Research Mohali,
  Knowledge City, Sector 81, S. A. S. Nagar,
  Manauli-140306, India
}

\date{\today}
\begin{abstract} \noindent Using large scale molecular dynamics
  simulations we study the orientational dynamics of a heated Janus
  particle which exhibits self-propulsion. The asymmetry in the
  microscopic interaction of the colloid with the solvent is
  implemented by choosing different wetting parameters for the two
  halves of the sphere. This choice leads to a different microscopic Kapitza
  resistance across the solid-fluid boundary of the two halves of the sphere, 
  and consequently a gradient in temperature is created across the
  poles of the sphere. It is this self-created temperature gradient
  which leads to a self-propulsion along the direction of the symmetry
  axis. In this article, we look at the orientational dynamics of such a system, as well as
  the subsequent enhancement of the translational diffusivity of the heated
  Janus colloid at late times. 
  The orientational correlation of the symmetry axis is measured from
  the simulation and provides a direct access to the rotational
  diffusion constant. The heating leads to an increase 
  in the rotational diffusivity of the colloid. 
  We quantify this increase in rotational diffusion
  $D_r$ against the temperature difference
  $\delta T \equiv T(R,0)-T(R,\pi)$ across the poles of the Janus
  sphere as well as the average surface temperature difference
  $\Delta T \equiv T(R)-T(\infty)$ from the ambient fluid. Since the
  rotational diffusion is determined by the complete flow field in the
  solvent, we illustrate that comparing $D_r$ against $\delta T$ is
  misleading and is better quantified when compared against
  $\Delta T$. The later quantification results in a data collapse for
  different choices of the microscopic interaction. The average
  propulsion velocity is also measured for different choices of the
  wetting parameter. The directionality of self-propulsion changes depending
  on the microscopic interaction. We show that whenever the attractive
  interaction of the colloid with the solvent is switched off, the
  phoretic mobility changes sign. Further, the propulsion velocity is
  zero for heating below a certain threshold value. This is also
  corroborated by the probability distribution of the angle between
  the displacement vector
  $\Delta \vec{r}(t) \equiv \vec{r}(t)-\vec{r}(0)$ and the symmetry
   axis. 
  Finally, we combine the measured propulsion
  velocity and the rotational diffusion time $\tau_r=1/2D_r$ to
  estimate the enhancement in the long time diffusion coefficient of
  the particle.

\end{abstract}

\pacs{...}

\maketitle

\section{Introduction}
Autonomous transport have gained a lot of attention from the
scientific community. A variety of artificial micro and nano swimmers have
been fabricated that exhibit different propulsion mechanisms.The
recent impetus to study the single particle and collective motions of
autonomous microswimmers is on one hand nourished by our interest to
have a better understanding of the functioning of the living world
down to the the microscopic scale, and on the other hand by the
diverse implications in technology, specifically the designing of
nanomachines \cite{Popescu:2010cq,Popescu:2011kf,
  Golestanian:2007hu,Jiang:2010el,Schoen:2006km,Chen:2000jv}. The key
ingredient in any self-phoretic motion of a colloid is an asymmetry in
the interaction with environment
\cite{Golestanian:2007hu,Popescu:2011kf}, the most common example
being the motion of a Janus particle in a thermal or a chemical
gradient or both. Typically, for a Janus particle, the two surfaces
have different chemical or thermal properties. For example,
self-diffusiophoretic motion has been achieved by coating one half of a
spherical particle with Platinum and immersing in a solvent containing
Hydrogen Peroxide\cite{Howse:2007ed}.  Similarly, self-thermophoretic
motion is obtained by coating half of a Polystyrene sphere with a
conducting material, such as Gold, and heating the composite particle
in the focus of a laser \cite{Jiang:2010el}.

The theoretical formulation of estimating the net drift velocity of a
microswimmer usually employs a hydrodynamic formulation where the slip
layer is assumed to be small and the resultant slip velocity in the
boundary layer is given by
$\vec{u}_s=\mu(\vec{r}_S) (\mathbf{I}-\vec{n} \vec{n})\cdot \nabla
\phi $,
where $\mu(\vec{r}_S)$ is the spatially dependent phoretic mobility,
$\vec{n}$ is the local normal to the surface and the choice of $\phi$
is dictated by the phoretic mechanism. For a diffusiophoretic
self-propulsion $\phi$ is the concentration field while $\phi$ is the
temperature field for thermophoretic self-propulsion. The drift
velocity is then given by averaging the negative surface slip velocity
$\vec{u}_s$ over the particle surface $\mathcal{S}$
\cite{Golestanian:2007hu}. However, the hydrodynamic model, by its
very definition coarse grains over the microscopic length scales and
the complex dynamical processes at the microscopic level are often
subsumed into the macroscopic transport coefficients such as the
phoretic mobility. On the other hand, in experiments often these
phenomena at the microscopic scale are not accessible. The gap is
readily bridged by numerical simulations. 
\cite{Yang2011a,Yang2013a,Yang2014,Kharazmi:2015ha,Kharazmi:2017ga}

In this article, using large scale molecular dynamics simulation,
we study the self-thermophoresis of a Janus particle, with focus on the 
orientational dynamics of heated Janus colloid.
Since the rectification of particle motion by its self-propulsion is limited by the rotational
diffusion of its symmetry axis and effectively enhances the long time translational
diffusion coefficient, a systematic investigation of the orientational
dynamics needs to be done. While the translational degrees of freedom
has been well studied, limited results exists for the dynamics of the
orientational degrees of freedom\cite{Yang2014,Yang2013a,Yang2011a,Bickel2014a}. 
We use a simple model system that was proposed earlier and is known to 
generate self-propulsion \cite{Schachoff2015,Kroy2016}. 
The asymmetry in the microscopic interaction is implemented through the 
modification of the strength of the attractive term in the $12-6$ 
Lennard-Jones potential via the wetting parameter $c_{\alpha \beta}$ 
(see \cref{sec:model_system} for more details). The model has several
advantages -- the attraction can be switched on and off by choosing a particular 
value of $c_{\alpha \beta}$ and consequently the direction of propulsion 
can be changed. Additionally, the wetting parameter can also be made a 
continuous function of the polar angle $\theta$, effectively 
reducing the area of the model gold cap which is heated. In the present 
scenario we make the simplest choice of choosing constant values of
 $c_{\alpha \beta}$ for the two different hemispheres of the spherical colloid.
 In our earlier work \cite{Kroy2016}, while the model system was used to look at the 
 temperature profile and the propulsion velocity generated as function of
 the heating of the colloid and to investigate the size of the slip layer around
 the heated Janus colloid, a systematic discussion on the orientational degrees of 
 freedom and the subsequent enhancement of the long time translational diffusivity was not done.
 Hence, we focus on the orientational dynamics of a heated Janus particle, with the particular aim
 to look at the dependency of the rotational diffusion coefficient on the heating
 of the colloid and the microscopic wetting parameter.  
 We summarize this section by pointing out that the main results of this work
 are contained in 
 \cref{fig:arrhenius_behavoir,fig:rotational_diffusion_constant_deltaT,fig:effective_diff}.

The rest of the paper is organised as follows. In \cref{sec:model_system}
we present our model system and the simulation details. The 
statics and dynamics of the symmetry axis is discussed in 
\cref{sec:orientational_dynamics}. The propulsion velocity of the colloid 
and the enhancement in the translational diffusion coefficient 
at late times with increased heating of the colloid is presented
in \cref{sec:propulsion_vel}. Finally, a brief conclusion and
outlook is presented in \cref{sec:conclusion}.

\section{Model System and Simulation Detail}
\label{sec:model_system}
Our simulation model consists 
of a nanoparticle immersed in a
Lennard-Jones solvent. The microscopic interactions between the
colloid and the solvent is given by the Lennard-Jones $12-6$
potential,
$U_{\alpha \beta}(r)=4 \epsilon \left[\left(\sigma/r \right)^{12}
  -c_{\alpha \beta}\left(\sigma/r\right)^6\right]$,
where $c_{\alpha \beta}$ is the wetting
parameter\cite{Schachoff2015,Kroy2016}. The nanoparticle is modelled as
a spherical cluster of Lennard-Jones particle, bound together by the
strong FENE potential $U(r)=-0.5\kappa R_0^2 \ln (1-(r/R_0)^2)$, with
$\kappa=30 \epsilon/\sigma$ and $R_0=1.5 \sigma$. Further, the Janus
particle is constructed from this spherical colloid by identifying an
outer shell of atoms in the upper hemisphere of the colloid as a model
Gold cap.  The wetting parameter, introduced above, can take different
values for the different types of pair interaction, namely
$c_{ss}$ for the solvent-solvent, $c_{gs}$ for the model Gold cap -solvent, $c_{ps}$
for the model Polystyrene half-solvent, $c_{pg}$ for Polystyrene-Gold and $c_{gp}$ for Gold-Polystyrene. 
The broken symmetry between the two
surfaces of the colloid is implemented by choosing different values
for the wetting parameter in the interaction with the solvent, which
effectively varies the minimum of the pair potential as
$(2\sigma^6/c_{\alpha \beta})^{1/6}$. Thus, for the value of
$c_{\alpha \beta}=2$, the minimum distance between the centers of a
particle in the colloid and the solvent is $\sigma$, whereas for
$c_{\alpha \beta}=0$ the soft attractive part of the potential is
completely lost. The value of $c_{\alpha \beta}$ in the
solvent-solvent interaction is kept constant at $c_{ss}=1.0$. A
variation in $c_{ss}$ is not important in the present discussion,
since it only changes the phase diagram of the bulk solvent. Additionally, 
the wetting parameters for Gold-Polystyrene and Polystyrene-Gold interaction were fixed 
at $c_{pg}=c_{gp}=1$. A typical simulation run consists of an equilibration 
phase in the NPT ensemble,
with a N\'ose--Hoover thermostat and barostat, at a temperature of
$T_0=0.75 \epsilon/k_{\rm B}$ and a thermodynamic pressure of
$p=0.01 \sigma^3/\epsilon$\cite{Chakraborty:2011kk}.  After the initial equilibration, 
for the rest of the simulation run, the system was evolved in a heating phase.
During the heating phase, the global thermostat was switched off and the
temperature $T_p$ of the gold cap was continuously controlled by a momentum
conserving velocity rescaling procedure. Using a similar rescaling
procedure, the fluid at the boundary of the simulation box was
thermostated at the temperature $T_0$. For each nanoparticle
temperature, at least three trajectories of $1.5 \times 10^7$ steps
with $\delta t =0.005\tau$ were simulated. All data was collected once
the system reached a steady state. Throughout this paper, the value of
a physical observable is obtained by averaging over these independent
trajectories and the error bars denote the standard deviation of the
mean. Further, mass, length, energy and time is measured in units of $m$,
$\sigma$, $\epsilon$ and $\tau=\sqrt{m\sigma^2/\epsilon}$.
\begin{figure}[!t]
\begin{tikzpicture}
\node at (-50pt, 60pt) {{\it \large (a)}};
\node at (85pt, 60pt) {{ \it \large (b)}};
\node[above,right] (img) at (-2.5,0)
{  \includegraphics[width=0.3\linewidth]{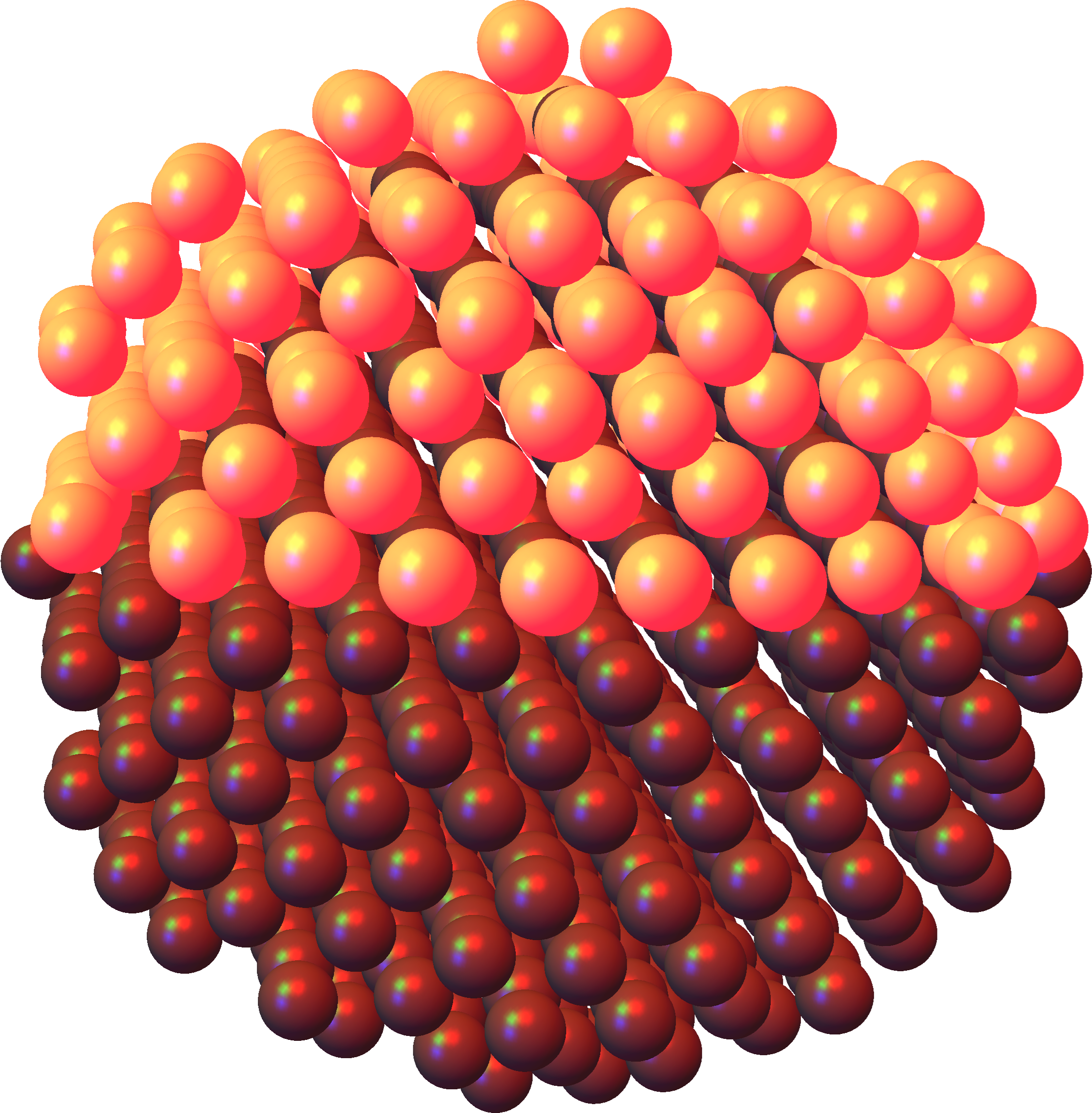}};
\node[above,right] (img) at (0.5,0)
{  \includegraphics[width=0.2\linewidth]{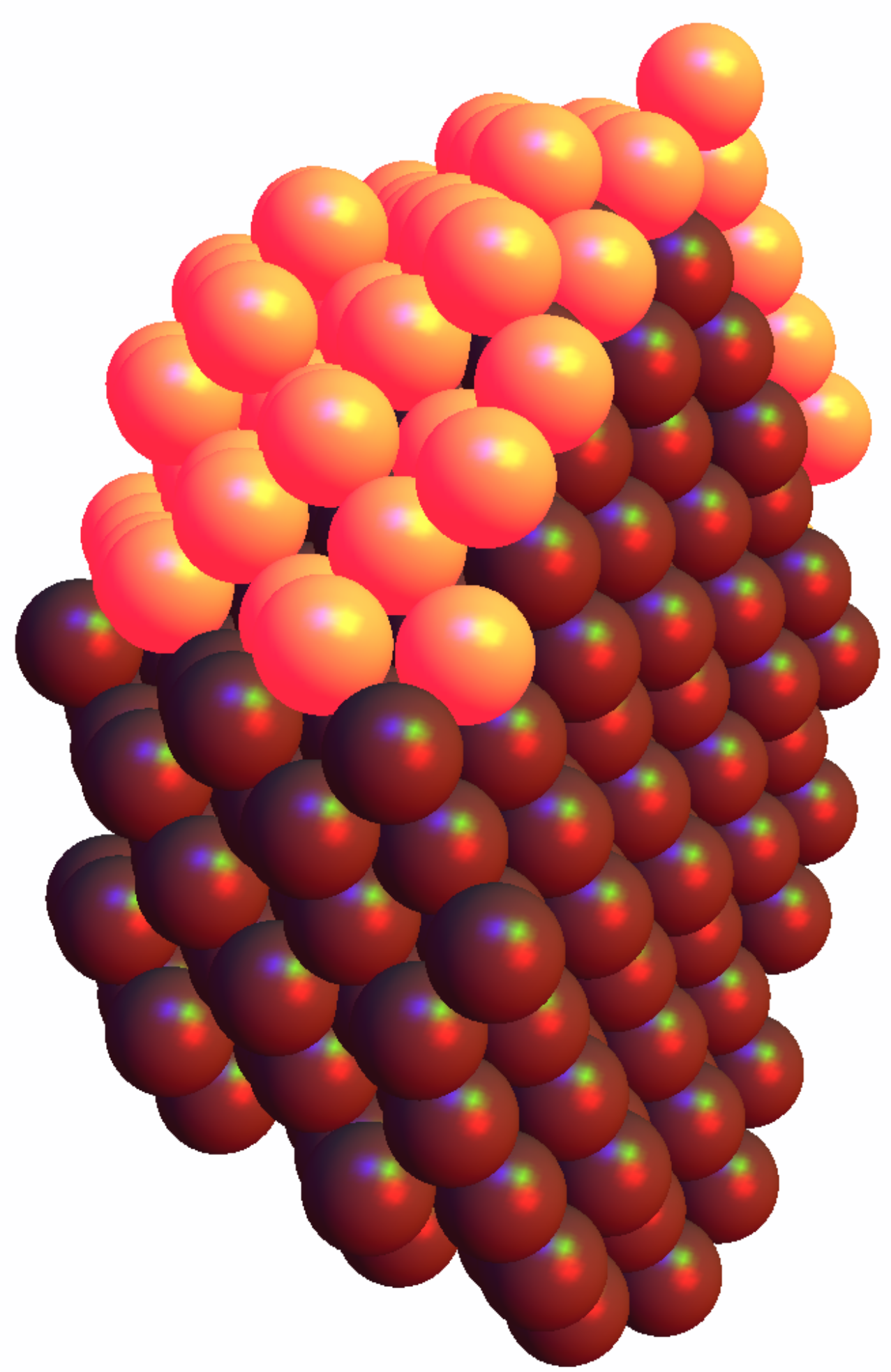}};
\node[above,right] (img) at (2.5,0)
{\includegraphics[width=0.48\linewidth]{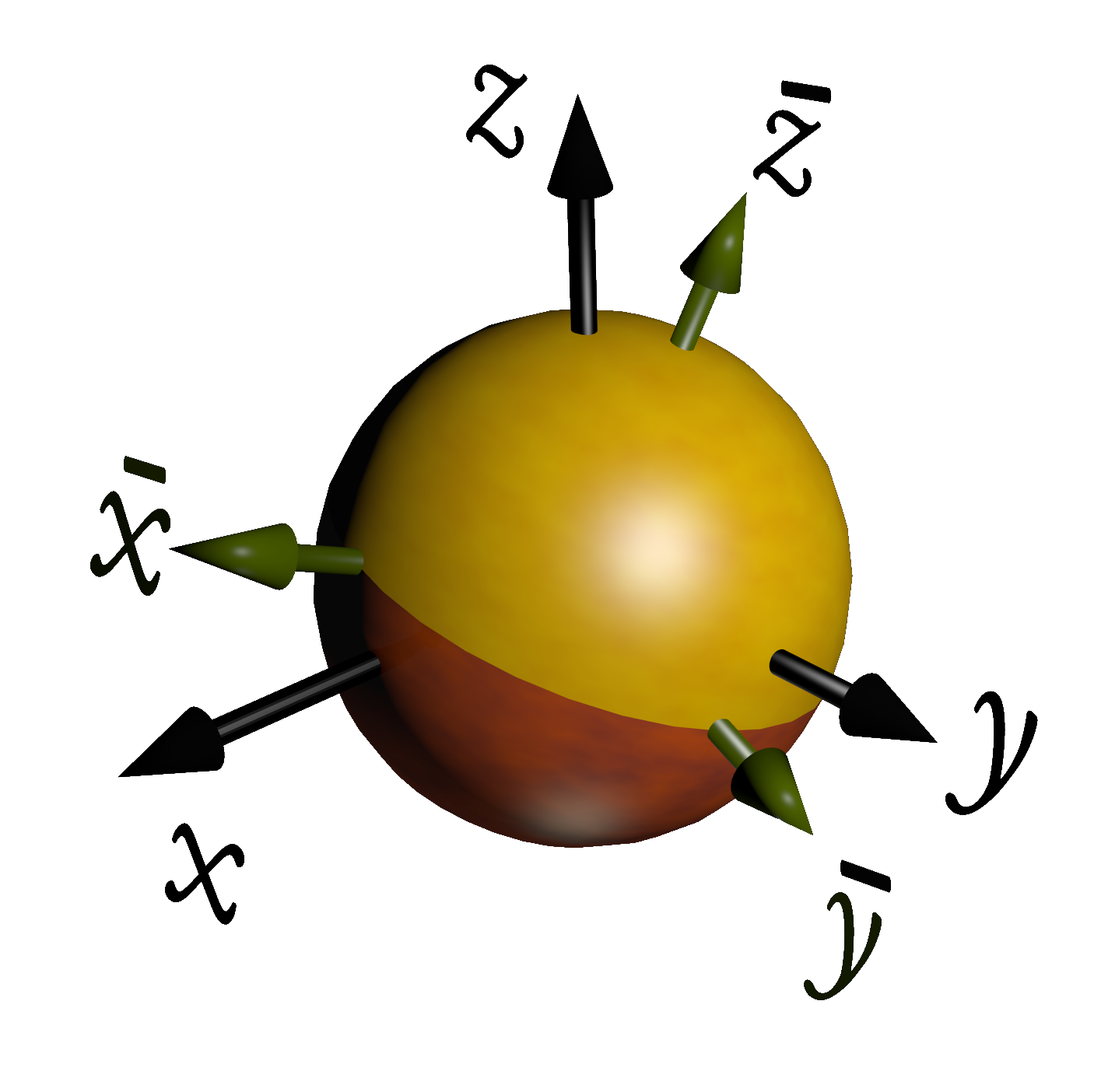}};
\end{tikzpicture}
\caption{\textbf{Figure (a)}: A snapshot of the Janus particle used
  from the simulations. First, the colloidal particle is constructed
  from the spherical cut of the initial FCC lattice of radius
  $5 \sigma$. The Janus particle in then constructed from this
  spherical cluster by identifying a layer of atoms of thickness
  $1.5 \sigma$ on the upper half of the sphere as a model gold cap
  (denoted by yellow) and rest as polystyrene. The interaction of the
  two species of atoms in the cluster with the solvent are different
  (see main text for details). \textbf{Figure (b)}: Schematic
  illustration of the body and the lab frame. The lab frame is fixed
  to the laboratory ($x,y,z$), whereas the body frame is fixed to
  colloid and rotates with the particle ($\bar{x},\bar{y},\bar{z}$). 
  Taken from Kroy \emph{et. al.}, Eur. Phys. J. Spec. Top., \textbf{225}, 2207 (2016)
  \cite{Kroy2016}, with kind permission of the European Physical Journal (EPJ).}
\label{schematic_illustration}
\end{figure}

A simple estimation of the heat diffusivity from the measured compressibility $\kappa_T$, 
and specific heat $c_p$ of the bulk solvent at the thermodynamic state of $(T_0,p)$ yields 
$D_T \equiv \kappa_T/\rho c_p \sim 2 \sigma^2/\tau$.  
On the other hand, the measured diffusivity of the Brownian
particle at the same thermodynamic state point gives us 
$D \equiv k_B T/6\pi\eta R \sim 0.003 \sigma^2/\tau$ ($R$ is
the radius of the particle and $\eta$ is the viscosity of the solvent). 
The large value of the ratio $D_T/D$ 
supports the claim of time-scale separation between heat propagation
and Brownian motion -- heat diffuses much faster than the colloidal diffusion.  
As a consequence of the resulting NESS, the
Brownian particle maintains a spatially varying and comoving
temperature profile $T(r)$. 
At the microscopic level, there exists an interfacial thermal resistance (Kapitza resistance)
across the solid-fluid boundary leading to a temperature discontinuity at the interface. \cite{Schachoff2015}
This temperature discontinuity depends on the parameters of interaction
between the solvent and colloid \cite{Vladkov:2006ema,Barrat:2003dx}. 
In the molecular dynamics simulations, we exploit this microscopic phenomenon 
at the solid-fluid interface to create a temperature gradient $\delta T$ between 
the north-south poles of the spherical colloid. \cite{Kroy2016} It is this gradient that 
generates an autonomous motion of the colloid.

\begin{figure}[!t]
  \centering
  \includegraphics[width=\linewidth]{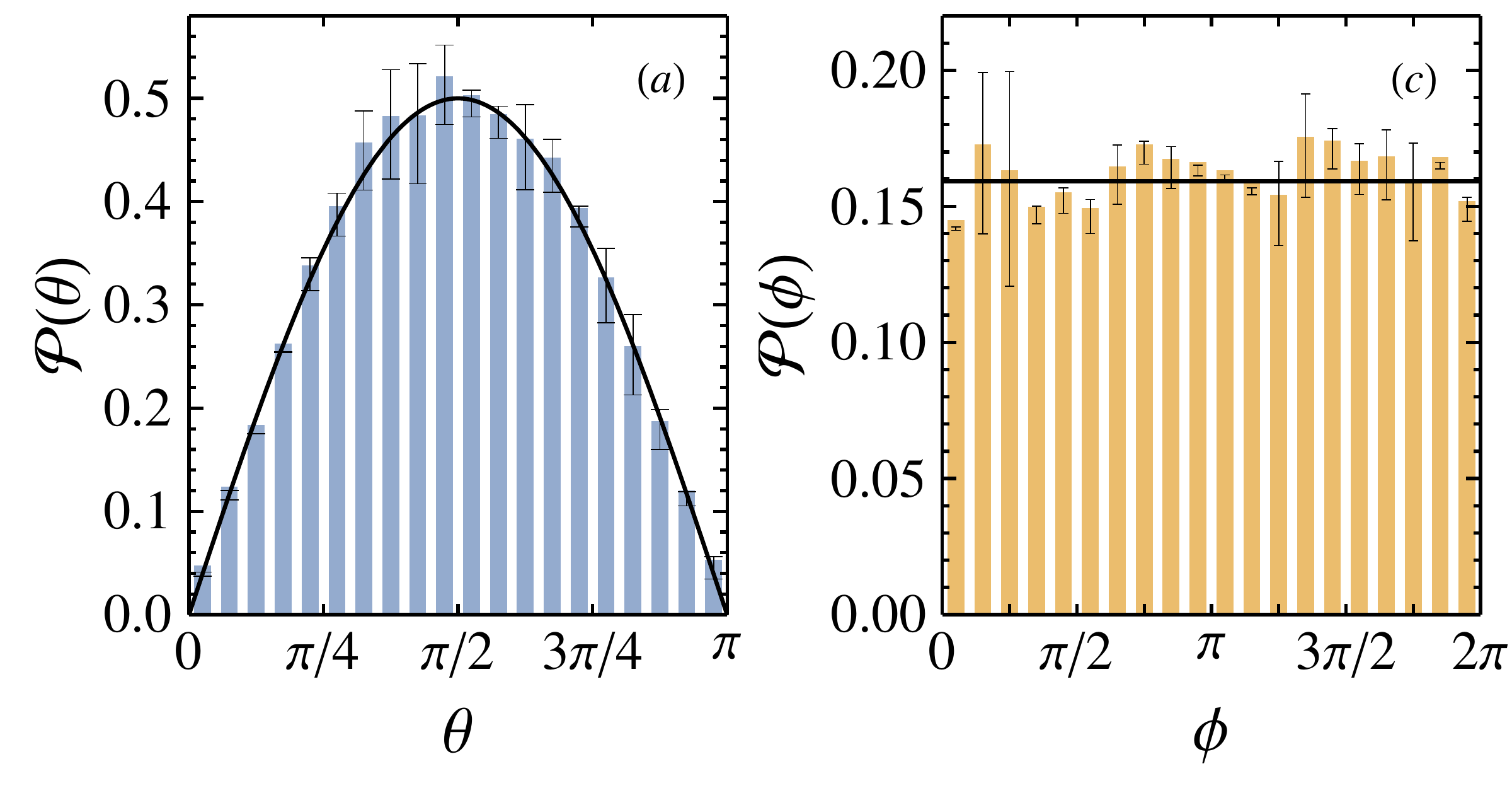}
  \caption{Distribution of the angle $\theta$ (figure (a)) and $\phi$
    (figure (b)), the symmetry axis and its orthogonal projection
    makes with the $z$-axis and the $x$ axis of the lab frame. The
    solid lines in the figures are plots of the theoretical prediction
    \cref{eq:theta_distribution} and \cref{eq:phi_dist}.The particle
    temperature is $\kb T_p/\epsilon=1.50$.}
\label{fig:theta_distribution}
\end{figure}

\section{Orientational Dynamics}
\label{sec:orientational_dynamics}

In order to look at the orientational dynamics of the heated Janus
particle, we recorded the unit vectors
$(\bar{\vec{x}},\bar{\vec{y}},\bar{\vec{z}})$ in the body frame of the
particle. The schematic illustration of this is shown in
\cref{schematic_illustration}~(b). Throughout this article the vector
$\bar{\vec{z}}$ as denoted in the figure corresponds to the symmetry
axis $\hat{\vec{n}}$. At the beginning of the simulation, both the
body and the lab frame coincide with each other. Accordingly, the unit
vectors in the body frame were constructed at the beginning of the
simulation by identifying the particles within the spherical cluster
which were furthest from the center of colloid, along the three
perpendicular directions. In the subsequent time steps, the unit
vectors were computed by noting the positions of these particles
relative to the center of mass of the colloid.

The hydrodynamic modeling of self-propulsion considers the motion
to be force and torque free, and the symmetry axis performs a free
diffusion on the surface of a sphere.
In the lab frame, the symmetry axis is given by
$\bar{\vec{z}}=(\sin \theta \cos \phi, \sin \theta \sin \phi, \cos
\theta)$,
where $\theta$ and $\phi$ are polar and azimuth angle, respectively,
that $\bar{\vec{z}}(t)$ makes with the lab frame. The joint
probability $p( \cos \theta, cos \phi, t)$ follows the
relation\cite{Teeffelen2010}
 \begin{equation}
 \label{eq:prob_dist_theory}
 p(\cos \theta, \cos \phi ,t)=\sum_{l,m}e^{-l(l+1)D_r t} Y_{lm}(\cos \theta)
 \end{equation}
 In the steady state, with $t \to \infty$ only the $l=0$ term survives
 and the stationary probability reads
 \begin{equation}
 \label{eq:prob_dist_stationary}
 p(\cos \theta, \cos \phi)=\frac{1}{4 \pi}
 \end{equation}
 Integrating out $\phi$ from the joint probability
 $p(\cos \theta, \cos \phi)$ and transforming to the variable $\theta$
 from $\cos \theta$ the probability $\mathcal{P}(\theta)$ reads
\begin{equation}
  \label{eq:theta_distribution}
\mathcal{P}(\theta)=\frac{1}{2} \sin \theta.
\end{equation}
A similar procedure to evaluate $\mathcal{P}(\phi)$ yeilds
\begin{equation}
\label{eq:phi_dist}
\mathcal{P}(\phi) =\frac{1}{2 \pi}.
\end{equation}

To check whether the microscopic model confirm
\cref{eq:theta_distribution} and \cref{eq:phi_dist}, we measured
$\theta$ and $\phi$ from the trajectories of $\bar{\vec{z}}(t)$ and
constructed the probability density function (PDF) from independent
configurations of the symmetry axis.  The measured PDF, plotted in
\cref{fig:theta_distribution} agrees
well with the theoretical prediction of \cref{eq:theta_distribution}
and \cref{eq:phi_dist}.

To look at the dynamics of the symmetry axis, we measure the 
orientational correlation time $\tau_r$ from the simulations.
An unbiased method to determine $\tau_r$ is to look into the correlation of the
unit vectors in the body frame. For a rigid spherical object in an
isothermal solvent, the orientation vector undergoes Brownian motion
on the surface of sphere. Due to the symmetry of the sphere, the
correlation of any orientation vector decays exponentially as:
\begin{equation}
\label{eq:orientational_corr}
C_{\bar{\vec{z}}}(t)\equiv  \langle \bar{\vec{z}}(t) \bar{\vec{z}}(0) \rangle=e^{-2D_r t}.
\end{equation}

\begin{figure}[!t]
  \centering
  \includegraphics[width=\linewidth]{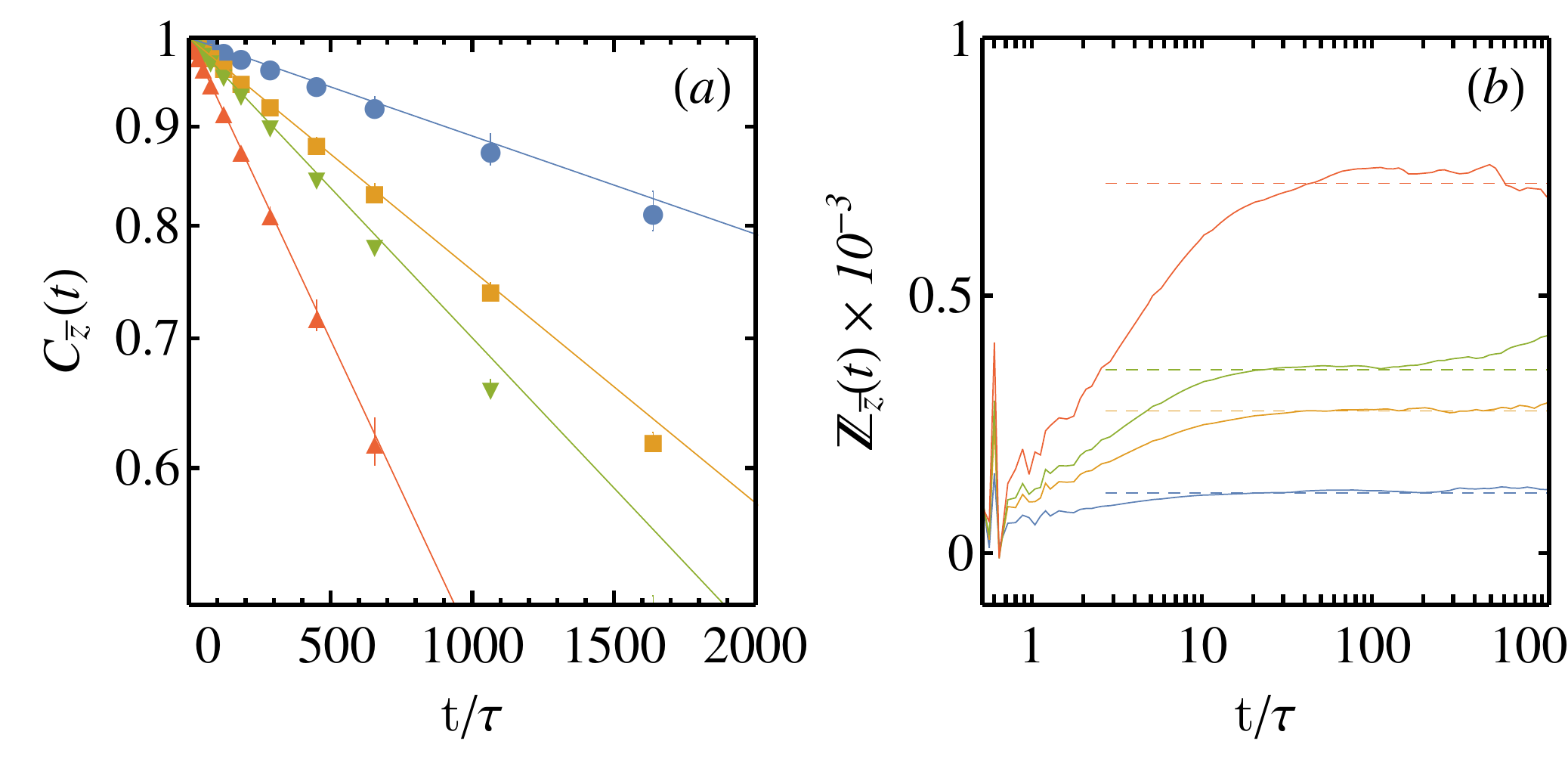}
  \caption{The dynamics of the unit vectors
    $\bar{\vec{z}})$ in the body frame is
    captured by their time correlation functions. \textbf{Figure (a)}:
    Correlation of the symmetry axis $\bar{\vec{z}}$ as a function of
    time for different heating of the colloid, $\kb T_p/\epsilon=0.8$
    ({\color{myblue} \large $\bullet$}), $1.30$ ({\color{myokker}
      $\blacksquare$}),$1.50$ ({\color{mygreen} $\blacktriangle$}) and
    $2.00$ ({\color{myorange} $\blacktriangledown$}). The solid lines
    are fit to the data using \cref{eq:orientational_corr}.
    \textbf{Figure (b)}: The numerical derivative defined by
    \cref{eq:orientation_corr_deriv} with
    $\bar{\vec{n}}=\bar{\vec{z}}$.  The dashed lines are the measured
    values of the rotational diffusion constant from a fit to the data
    using the functional form of \cref{eq:orientational_corr}. The
    colors indicate different temperature of the Janus particle as
    detailed in figure (a).  
  }
\label{fig:ncap_xcap_corr}
\end{figure}

This provides us with a direct route to measure the rotational
relaxation time $\tau_r=1/2D_r$.
To ensure that the numerical fit gives us the correct value of the
rotational diffusion constant, we also numerically evaluated the
function $\mathbb{Z}_{\bar{\vec{n}}}(t)$ defined as
\begin{equation}
  \label{eq:orientation_corr_deriv}
\mathbb{Z}_{\bar{\vec{n}}}(t)=-\frac{\dot{C}_{\bar{\vec{n}}}(t)}{C_{\bar{\vec{n}}}(t)},
\end{equation}
where $\bar{\vec{n}}$ denote either of the unit vectors in the body
frame. Using \cref{eq:orientational_corr}, the function
$\mathbb{Z}_{\bar{\vec{n}}}(t)$ at late time should saturate to $D_r$
(see \cref{fig:ncap_xcap_corr} (b)). 

\begin{figure}[!t]
  \centering
  \includegraphics[width=\linewidth]{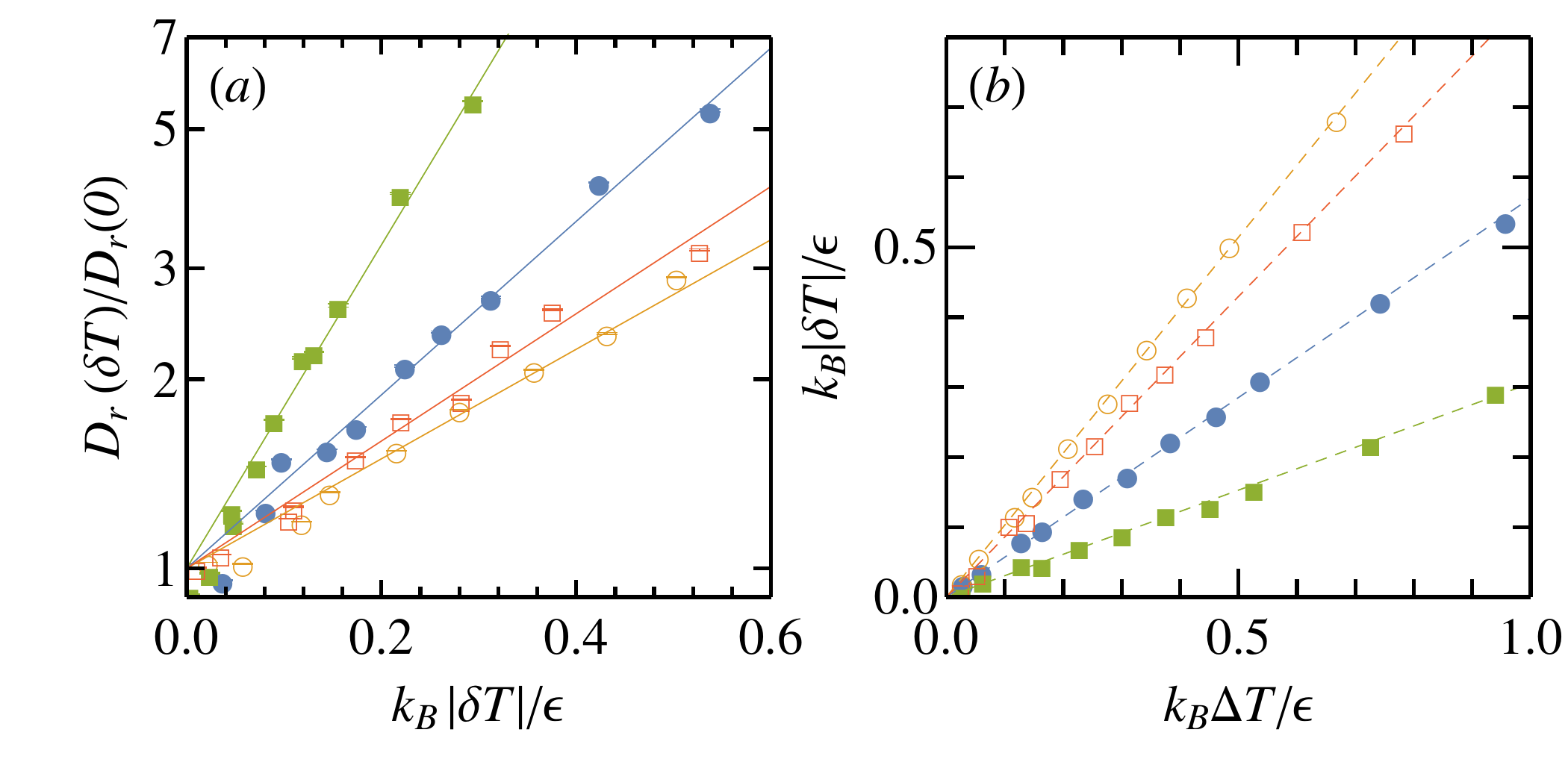}
 \caption{ (a):Plot of the normalized rotational
diffusion coefficient $D_r/D_r(0)$ as function of the heating of the
colloid for different choices of the wetting parameter:
 $c_{gs}=2, c_{ps}=1$ ({\color{myblue} \large $\bullet$});
 $c_{gs}=1,c_{ps}=2$ ({\color{mygreen} $\blacksquare$});
 $c_{gs}=2,c_{ps}=0$ ({\color{myokker}{\Large $\circ$}}); 
 and $c_{gs}=0,c_{ps}=2$ ({\color{myorange} $\square$}). 
 (b) Plot of the temperature difference $\delta T\equiv T(R,0)-T(R,\pi)$
  across the poles of the heated Janus particle against the incremental temperature
  difference from the ambient fluid $\Delta T=T(R)-T(\infty)$. The symbols indicate
  difference choices of the wetting parameters as noted above. }
\hfill
\label{fig:arrhenius_behavoir}
\end{figure}

\begin{figure}[!t]
  \centering
  \includegraphics[width=0.8\linewidth]{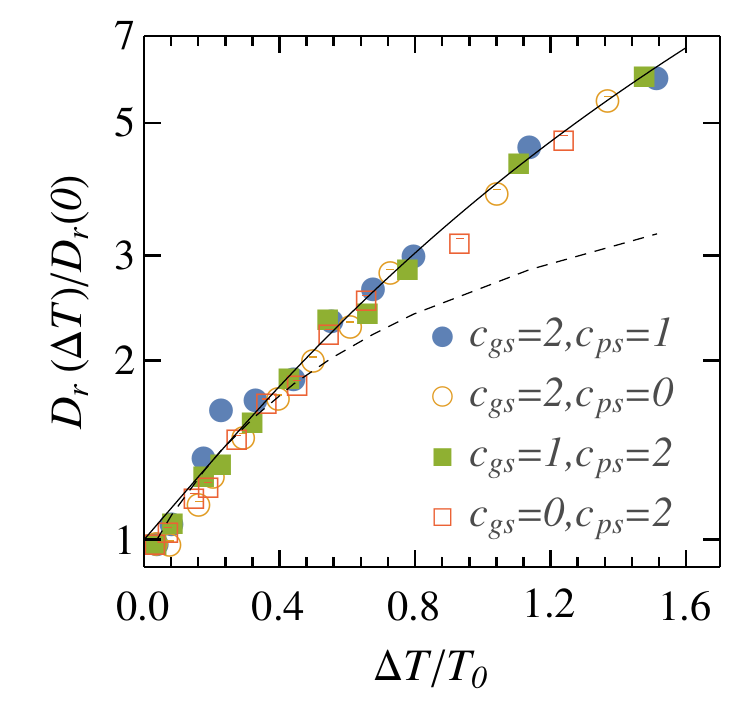}
\hfill
\caption{
  Plot of the normalized rotational diffusion constant as a function
  of the temperature difference $\Delta T \equiv T(R)-T(\infty)$ in
  the fluid on the particle surface and the ambient fluid far away
  from the Janus colloid. Different symbols indicate different choices
  of the wetting parameters as indicated in the legend. We obtain a data
  collapse for different choices of the wetting parameters. The solid
  line is guide to the eye and the dashed line is the theoretical
  prediction from hot Brownian motion using
  \cref{eq:THBM_rot,eq:zeta_final,eq:lj_viscosity}. 
}
\label{fig:rotational_diffusion_constant_deltaT}
\end{figure}

To further quantify the contribution of the rotational dynamics
towards the long time diffusivity, we measured $D_r$ for different
heating's of the Janus particle and different choices of the wetting
parameter $c_{\alpha \beta}$.
The rotational Brownian motion is a hydrodynamic
phenomena which requires looking at the fluid flow generated in the
whole system. Therefore, a quantification of $D_r$ with heating of the
colloid entails a subtlety of comparing against the local temperature
difference $\delta T \equiv T(R,0)-T(R,\pi)$ or
$\Delta T \equiv T(R)-T(\infty)$,the average temperature increment on the surface of the colloid above the
ambient temperature $T(\infty)\equiv T_0$. The former can lead to a wrong interpretation of
the data as illustrated in \cref{fig:arrhenius_behavoir}~(a)
and (b).
Even though the magnitudes are same for the complementary
choices of the wetting parameters (such as $c_{gs}=2, c_{ps}=1$ and
$c_{gs}=1, c_{ps}=2$ ), $D_r$ plotted against $\delta T$ shows a
counter-intuitive trend as expected from the behaviour of $\delta T$ as
a function of $T_P$ (see
\cref{fig:arrhenius_behavoir}~(b)). Note that $T_P$ is synonymous with $\Delta T$.
Had the rotational diffusion being a local phenomena, we would have expected
that for increased values of $\delta T$, the particle would rotate
faster and the rotational diffusion would (since
$D_r \sim \tau_r^{-1}$) also increase. Thus, we observe that for the
choice of $c_{gs}=1,c_{ps}=2$ (filled squares in
\cref{fig:rotational_diffusion_constant_deltaT}~(b)) the increase in
$\delta T$ is least, and therefore, we expect that the increase in rotational
diffusion should be least. However, looking at
\cref{fig:rotational_diffusion_constant_deltaT}~(a) we see that the
increase in $D_r$ is maximum (filled squares). A similar observation
is also observed for the cases when $c_{gs}=2,c_{ps}=0$ and
$c_{gs}=2,c_{ps}=0$. As a passing remark we note that $D_r$ follows an
Arrehnius behavior when plotted against $\delta T$:
 \begin{equation}
   \label{eq:Drot_arrehnius}
 D_r(\delta T) = D_r(0)~ e^{\,\lambda \,\delta T}.
 \end{equation}
 The value of $\lambda$ varies for different choices of the wetting
 parameter, with $\lambda \approx 6$ is the maximum value for
 $c_{gs}=1$ and $c_{ps}=2$ and $\lambda \approx 2$ for $c_{gs}=0$ and
 $c_{ps}=2$ as well as for $c_{gs}=2$ and $c_{ps}=0$

 To properly quantify the effect of heating on rotational diffusion of
 the heated Janus particle we take a different approach. 
 Since rotational 
 diffusion of an isothermal as well as heated colloid \cite{Rings2012a} 
 is known to be a hydrodynamic phenomena, we quantify the effect of heating on $D_r$
 by comparing it against $\Delta T$. To this end, 
 we look at the temperature profile in the solvent by averaging the
 kinetic energy in spherical shells of thickness $0.5 \sigma$. The
 resulting radial temperature field is fitted to the solution of the
 Fourier's heat equation with a temperature dependent thermal
 conductivity\cite{Chakraborty:2011kk}
 \begin{equation}
   \label{eq:temp_prf}
T(r)=T_0\left(1+\frac{\Delta T}{T_0}\right)^{R/r}.
\end{equation}
In order to take care of the periodic boundary condition imposed on
the simulation box, the fit was used to determine the incremental
temperature $\Delta T$ and the ambient temperature $T_0$. The value of
$\Delta T$ and $T_0$ completely specifies the far away temperature
field in the system and when $D_r$ is plotted against the ratio
$\Delta T/T_0$ we observe a data collapse for all the choices of the
wetting parameters. To proceed further and quantify $D_r$ in terms of
$\Delta T/T_0$, we ignore the asymmetry in the temperature and the
viscosity profile and use the framework of rotational hot Brownian
motion\cite{Rings2012a}. This approximation should be valid for small
heating of the Janus particle, so that the non-linear effects due to
the asymmetry does not appear \cite{Auschra2017}. The rotational
diffusion coefficient of the heated particle obeys the generalized
Stokes-Einstein relation
 \begin{equation}
\label{eq:generalized_ser}
   D_r=\kb T^{\rm r}_{\rm HBM}/\zeta^{\rm r}_{\rm HBM},
 \end{equation}
 where $T^{\rm r}_{\rm HBM}$ and $\zeta^{\rm r}_{\rm HBM}$ are the
 effective temperature and friction coefficient of the heated
 particle. Using fluctuating hydrodynamics, the effective parameters
 can be evaluated from the steady state temperature profile $T(r)$ and
 the viscosity profile $\eta(r)$
 \cite{Rings2012a,Brilliantov1991}. For convenience we reproduce the
 expressions for $T^{\rm r}_{\rm HBM}$ and $\zeta^{\rm r}_{\rm HBM}$
 that reads as:
 \begin{equation}
   \label{eq:THBM_rot}
T^{\rm r}_{\rm HBM}=\frac{\int_R^\infty T(r)\eta^{-1}(r)r^{-4}\,\upd r}
  {\int_R^\infty \eta^{-1}(r)r^{-4}\,\upd r}\,
 \end{equation}
and 
\begin{equation}
  \label{eq:zeta_final}
  (\zeta^{r}_{\rm HBM})^{-1} = \frac{3}{{8\pi}}
  \int_R^\infty\!\!\frac{1}{\eta(r) r^{4}}\,\upd r \,.
\end{equation}

Under isothermal conditions, with $\Delta T=0$, the effective
temperature and the effective friction coefficient becomes
$T^{\rm r}_{\rm HBM}=T_0$ and
$\zeta^{\rm r}_{\rm HBM}=8 \pi \eta_0 R^3$, where $T_0$ is ambient fluid
temperature and $\eta_0$ is the bulk viscosity of the solvent. 

In a Lennard--Jones system the temperature dependence of the viscosity
is given by\cite{Chakraborty:2011kk}
\begin{equation}
  \label{eq:lj_viscosity}
\log \left[\frac{\eta(T)}{\eta_\infty}\right]={A \over T^4}.
\end{equation}
Using the radial temperature profile from \cref{eq:temp_prf}, we
transform the temperature dependence of the viscosity to a radial
dependence, and use \cref{eq:THBM_rot} and \cref{eq:zeta_final} to
numerically calculate the rotational diffusion constant for different
values of $\Delta T/T_0$. The numerical data is shown by the dashed line in
\cref{fig:rotational_diffusion_constant_deltaT}, which agrees
well for small values of $\Delta T/T_0$. A phenomenological
fit using a quadratic polynomial
\begin{equation}
	{D_r(\Delta T) \over D_r(0)} \approx 1+ \frac{3}{2} \left( \frac{\Delta T}{T_0} \right)+\frac{5}{4} \left( \frac{\Delta T}{T_0}\right)^2
\end{equation}
gives reasonable fit to the data for
the whole range of  $\Delta T/T_0$ investigated in the current work
(the solid line in \cref{fig:rotational_diffusion_constant_deltaT}).

\section{Enhancement in translational diffusivity}
\label{sec:propulsion_vel}
We next turn our attention to the enhancement of the late time diffusivity
in the translation motion of the heated Janus particle. 
Ordinarily, for a spherical particle, the
translational and rotational degrees of freedom are independent of
each other, even in the scenario of symmetric heat absorption of the
colloid. However, in self-phoretic motions, the different modes of
motion become coupled, resulting in an enhanced diffusion of the
colloid at late times. In the lab-frame, the Langevin equation in the
overdamped limit for a self-propelled particle reads
\begin{equation}
  \label{eq:langevin_equation}
\dot{\vec{r}}=V_p \bar{\vec{z}}+ \pmb{\xi}
\end{equation}
with
$\langle \xi_i(t) \xi_j(t') \rangle =2 D \delta(t-t') \delta_{ij}$ and
$\langle \bar{\vec{z}}(t) \bar{\vec{z}}(t') \rangle=e^{-D_r|t-t'|}
$.

\begin{figure}[!t]
  \centering
  \includegraphics[width=0.8\linewidth]{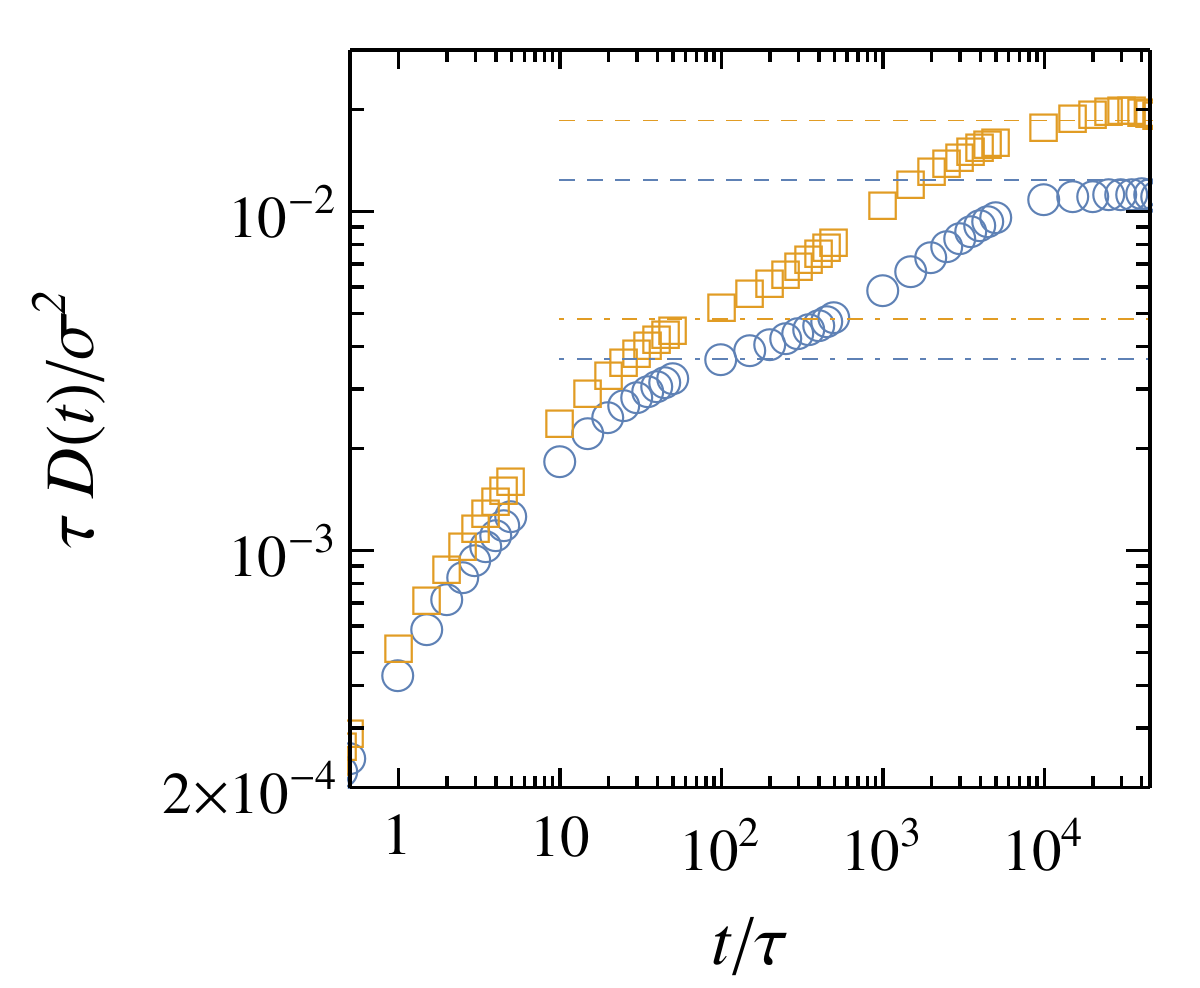}
\hfill
\caption{Time dependent diffusivity of a heated Janus colloid for
 temperatures $\kb T_P/\epsilon=1.5$ (open circles) and $2.0$ (open squares). 
 The dot-dashed lines corresponds to the translational diffusion coefficient $\DHBM$
 due to hot Brownian motion measured from the mean-square displacement 
 in the $\bar{\mathbf{x}}-\bar{\mathbf{y}}$ plane. The dashed line corresponds
 to the value $\DHBM+V_p^2 \tau_r/3$ with values of $\tau_r$ and 
 $V_p$ measured from the simulations.}
\label{fig:diffusivity}
\end{figure}

The mean-square displacement (MSD) $\langle \mathbf{r}^2(t)\rangle$
take the form,
\begin{equation}
  \label{eq:msd_lab}
  \langle \mathbf{r}^2(\Delta t)\rangle= 6D \Delta t + 2 V_p^2
    \tau_r^2 \left[ \frac{ \Delta t}{\tau_r} 
     + e^{-2 \Delta t/\tau_r}-1\right],
\end{equation}
where $\mathbf{V}_p$ is the net propulsion velocity of the colloid,
$D$ is the diffusion constant in absence of any broken symmetry and
$\tau_r$ is inverse of twice the rotational diffusion constant. 
In a frame of reference rotating with the colloid, the motions along
and perpendicular to the direction of propulsion are
decoupled. Without loss of generality, we assume that the $z$-axis of
the rotating frame is oriented along this direction (this also
corresponds to the symmetry axis of the capped particle), and the MSD
is given by
\begin{eqnarray}  \label{msd_body_frame} 
 \nonumber
  \langle \bar{x}^2(\Delta t)\rangle \equiv 
  \langle \bar{y}^2(\Delta t) \rangle = 2 D \Delta t \\
  \langle \bar{z}^2(t) \rangle = 2D \Delta t + V_p^2 \Delta t^2
\end{eqnarray}

At late times, the MSD in the lab frame from
\cref{eq:langevin_equation} shows an increased diffusivity $D_{\rm
  eff}$ given by
\begin{equation}
  \label{eq:diffusivity_late_times}
D_{\rm eff}=\DHBM+\frac{V_p^2 \tau_r}{3}=\DHBM (1+\chi),
\end{equation}
with $\chi=V_p^2 \tau_r/3\DHBM$.
The first term in \cref{eq:diffusivity_late_times} corresponds to the enhanced diffusivity
of the heated particle due to its hot Brownian motion \cite{Chakraborty:2011kk,Rings2012a} 
while the second term has two competing physical mechanisms - an increase in
the net propulsion velocity with heating and a decrease in 
orientational correlation time with heating. 

As a first check we wanted to validate \cref{eq:diffusivity_late_times} 
from the measured time dependent diffusivity of the heated Janus 
particle.This validation is shown in \cref{fig:diffusivity}, where we 
compare the long time diffusion coefficient with the predicted value 
in \cref{eq:diffusivity_late_times} and find excellent agreement of the 
predicted values with the measured diffusion
coefficient. 

The propulsion velocity is directly measured from the lab-frame particle 
velocity and projecting it in the body frame every time step. 
\cite{Chakrabarty:2013kw}
To quantify the propulsion velocity $V_p$ with the heating of the colloid, 
we choose to do it against both $\delta T$ as well as $\Delta T$, separately.
However, in order to make a consistent estimate of the enhancement factor in 
the late time translational diffusion coefficient we use the later quantification  
against $\Delta T/T_0$.\cite{Bickel2013a} It should be noted that since the
hydrodynamic modelling of self-phoresis is a boundary value problem, this
quantification of $V_p$ against $\Delta T/T_0$ is sufficient. We still choose to
use $\delta T$ primarily to show that the phoretic mobility changes sign whenever
the attractive interaction in either of the hemispheres is switched off. 

\begin{figure}[!t]
  \centering
  \includegraphics[width=\linewidth]{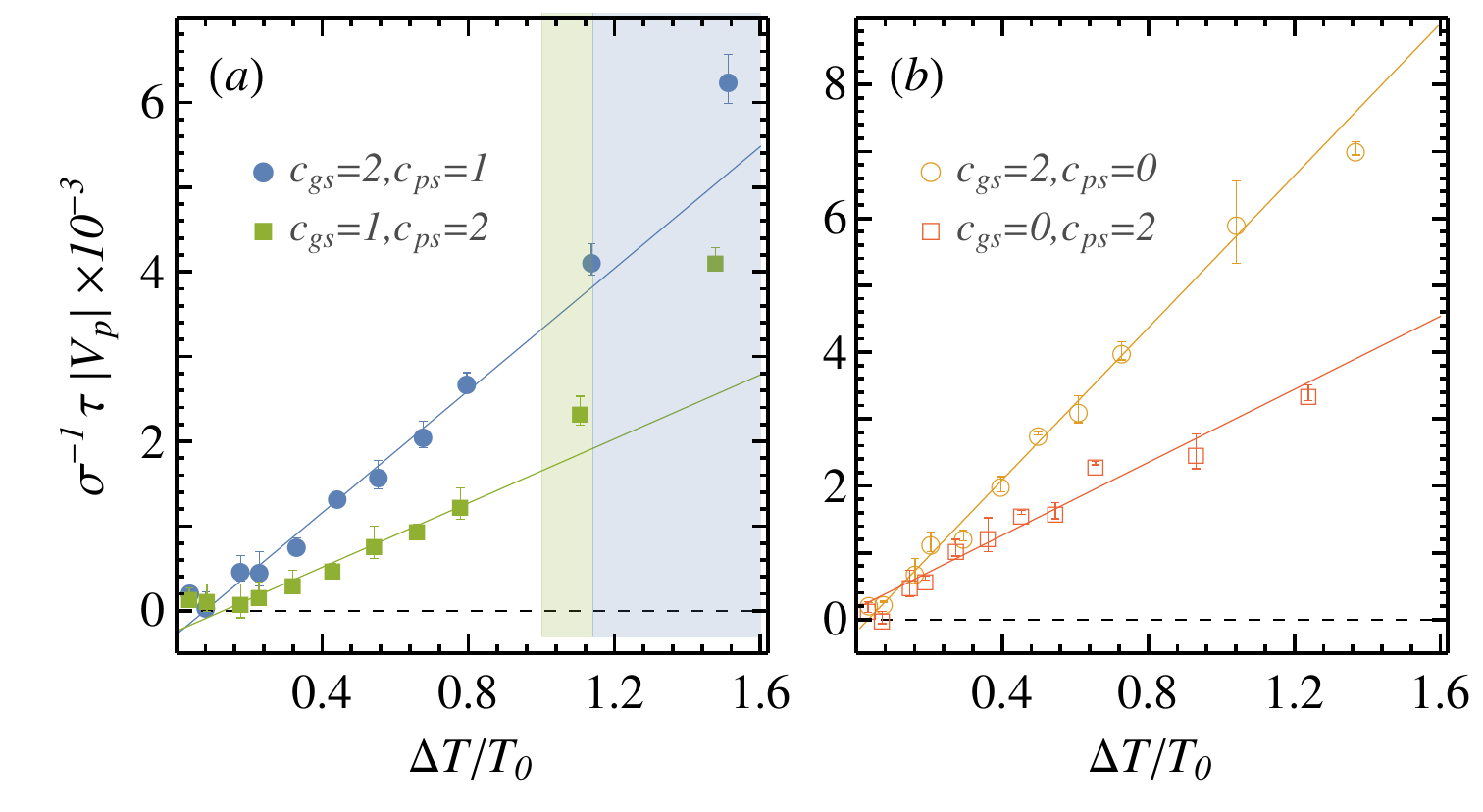}
\hfill
\caption{Magnitude of the propulsion velocity $V_p$ as function of the
  average fluid temperature difference $\Delta T /T_0$ on
  the surface of the colloid above the ambient temperature $_0$ 
  for different combinations of the wetting
  parameters: $c_{gs}=2,c_{ps}=1$ (filled circles in figure (a)), $c_{gs}=2,c_{ps}=0$
  (filled squares in figure (a)), $c_{gs}=1,c_{ps}=2$ (empty circles in figure (b)) and
  $c_{gs}=0,c_{ps}=2$ (empty circles in figure (b)). The solid lines are linear fit to
  the data, whereas the dashed lines denote the value $V_p=0$ clearly
  indicating that for small heating there no appreciable propulsion
  velocity.The shaded region in figure (a) indicate the
  non-linear regime.}
\label{fig:propulsion_vel}
\end{figure}

The measured propulsion velocities reveal that irrespective of the choice of
$c_{\alpha \beta}$, there always exists a threshold in heating below
which there no appreciable propulsion of the Janus particle (see \cref{fig:propulsion_vel})
and the motion is purely diffusive.
For higher heating's of the colloid, a net
propulsion in the direction of the symmetry axis emerges. This is also
corroborated by the measured time 
dependent diffusion coefficient.\cite{Kroy2016}

\begin{figure}[!t]
  \centering
  \includegraphics[width=\linewidth]{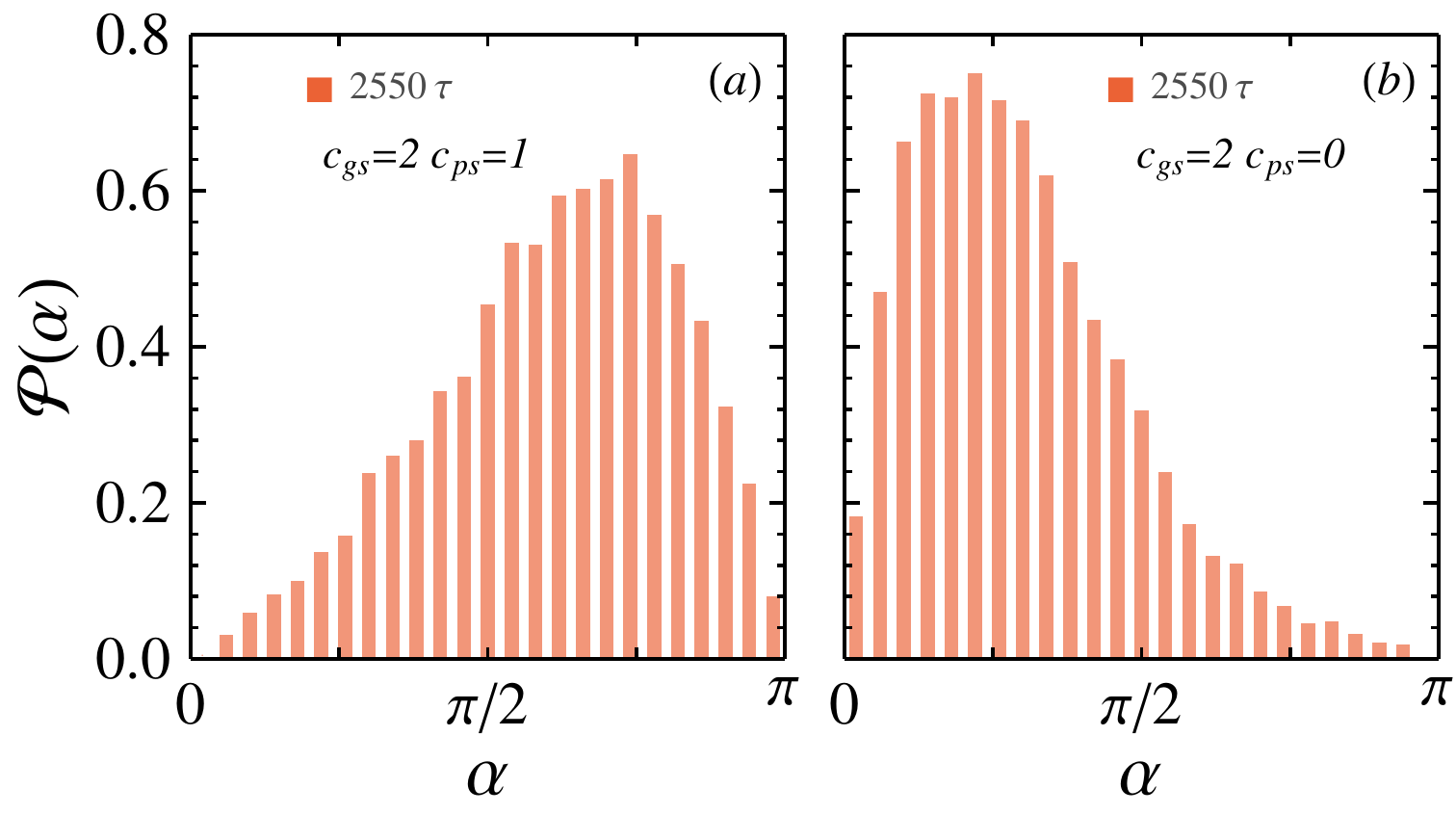}
  \caption{Distribution of the angle $\alpha$ between the symmetry
    axis $\bar{\vec{z}}(0)$ and the displacement vector
    $\Delta \vec{r}(t)$ for particle temperature
    $\kb T_P/\epsilon=1.20$ and wetting parameters $c_{gs}=2,c_{ps}=1$
    (figures (a) --(c)) and $c_{gs}=2,c_{ps}=0$ (figures (d)
    --(f)). In both the choices $\delta T >0 $ but the particle
    displacement changes as the sign of the phoretic mobility changes.
  }
\label{fig:alpha_dist_cg2cp1_cg2cp0}
\end{figure}

To explore
this further, we look closely at the trajectory of the particle and
calculate the angle $\alpha$ between the displacement vector
$\Delta \vec{r}(t)\equiv\vec{r}(t)-\vec{r}(0)$ and the symmetry axis
$\bar{\vec{z}}(0)$ for values of $t$ which are less than the
rotational relaxation time of the particle. In the absence of a directed motion, 
the rotational symmetry of the
system is still preserved and therefore in the body frame the unit
vector $\Delta \hat{\vec{r}}=\Delta \vec{r}/|\Delta \vec{r}|$ rotates
uniformly on a unit sphere. Consequently, the angle between the
displacement vector $\Delta \vec{r}$ and the symmetry axis would
follow a probability distribution given by
\cref{eq:theta_distribution}. Although, we do not show this,
it is indeed observed from the simulations, 
with the measured probability distribution of 
$\alpha$ following \cref{eq:theta_distribution}.
When a directed motion emerges for
higher heating of the colloid, the rotational symmetry is broken and
the displacement vector would be oriented either towards or opposite
to the symmetry axis $\bar{\vec{z}}$, depending on the sign of the
phoretic mobility. Accordingly, for a positive phoretic mobility the
motion of the particle is along the symmetry axis when $\delta T <0$
and opposite to the symmetry axis when $\delta T >0$. Similarly, for a
negative phoretic mobility the motion of the particle is along the
symmetry axis when $\delta T >0$ and opposite when $\delta T<0$. We
summarise these in the following:
\begin{align}
\nonumber
  \mu >0 \; \; \;\;
\begin{cases*}
\; \; \; \delta T > 0 & \text{propulsion opposite $\bar{\vec{z}}$}\\
\; \; \; \delta T < 0 & \text{propulsion along $\bar{\vec{z}}$}\\
\end{cases*}
\\
  \mu <0 \; \; \;\;
\begin{cases}
\; \; \; \delta T > 0 & \text{propulsion along $\bar{\vec{z}}$}\\
\; \; \; \delta T < 0 & \text{propulsion opposite $\bar{\vec{z}}$}\\
\end{cases}
\end{align}
Since the directionality depends on
the sign of the phoretic mobility, we looked at the distribution of
the angle $\alpha$ for different choices of the wetting parameters,
the particular cases with $c_{gs}=2,c_{ps}=1$ and $c_{gs}=2,c_{ps}=0$
are illustrated in \cref{fig:alpha_dist_cg2cp1_cg2cp0}. In the former
scenario, the displacements at late times happen opposite to the
symmetry axis, whereas when the attractive interaction is switched off
the displacements happen along the direction of the symmetry axis.
Although not shown in the plot, we observe a similar behavior when
$c_{gs}=1,c_{ps}=2$ and $c_{gs}=0,c_{ps}=2$. We therefore conclude
that in presence of an attractive interaction the average phoretic
mobility $\mu$ is positive and becomes negative whenever one surface
has a repulsive interaction. \cite{Luesebrink2011} This dependence on
the microscopic interaction raises the interesting fact that for a
particular choice of the wetting parameters would make phoretic
mobility zero and consequently, a dynamic switching of the wetting
parameters in the experiments can result in a dynamical switching
on/off of the transport, leaving hot Brownian transport unaffected.

Beyond the threshold value in heating, $V_p$ increases linearly in
accordance with linear response. Additionally, we also observed that a
non-linear regime in the propulsion velocity is achieved beyond an
upper cutoff whenever the combination of the wetting parameter is such
that the two halves of the Janus particle interacts with the solvent
either strongly or weakly. In contrast, whenever either surface has a
purely repulsive interaction, the propulsion velocity increases
linearly throughout the range of heating that has been
investigated.

\begin{figure}[!t]
  \centering
  \includegraphics[width=\linewidth]{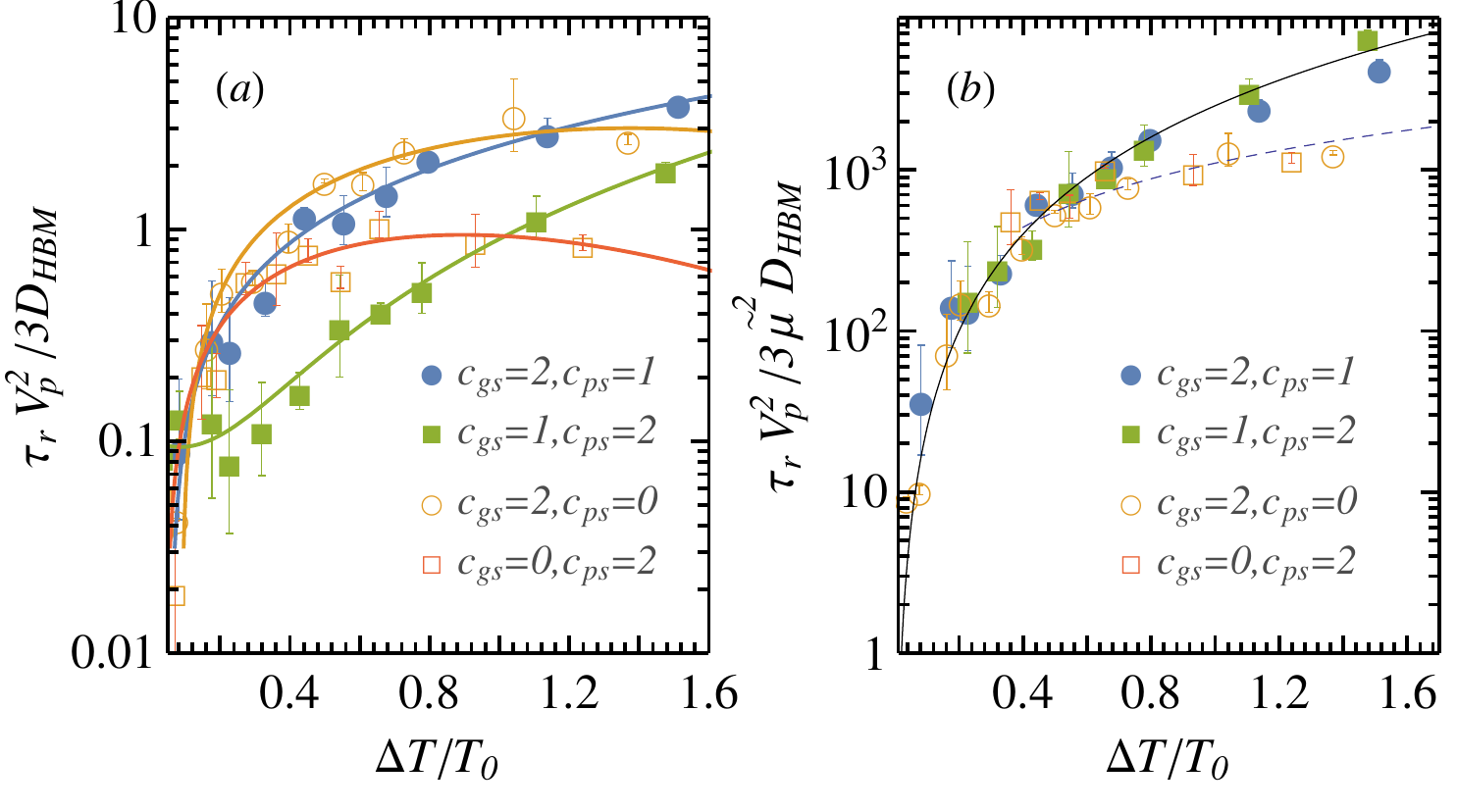}
\hfill
\caption{Plot of $\tau_r V^2_p$ (figure (a)) and the enhancement
  factor $\tau_r V^2_p/2D$ (figure (b)) against $\Delta T$ for
  different choices of the wetting parameters : $c_{gs}=2, c_{ps}=1$
  ({\color{myblue} \large $\bullet$}); $c_{gs}=2,c_{ps}=0$
  ({\color{myokker}{\Large $\circ$}});$c_{gs}=1,c_{ps}=2$
  ({\color{mygreen} $\blacksquare$}) and $c_{gs}=0,c_{ps}=2$
  ({\color{myorange} $\square$}). The lines are guide to the eye. }
\label{fig:effective_diff}
\end{figure}

Finally, we are in position to look at the term $V_p^2 \tau$ which
contributes to the long time diffusivity $D_{\rm eff}$. Since the
effective translational diffusivity is determined by the complete flow
field in the fluid\cite{Rings:2011gj,Chakraborty:2011kk,Rings2012a},
we choose to compare the augmenting term against $\Delta T/T_0$. A plot of the
term $V_p^2 \tau_r$ against $\Delta T/T_0$ for different choices of the
wetting parameters is shown in \cref{fig:effective_diff}~(a). The 
enhanced translational diffusion coefficient at late times can reach almost $4$
times the value of $\DHBM$- the value in absence of self-propulsion.
Comparing \cref{fig:effective_diff} with that of
\cref{fig:propulsion_vel}, we observe that whenever the propulsion is
in the linear response regime, which happens when the attractive
interaction with the solvent is switched off, the term $\tau_r V^2_p$
and therefore the enhancement factor approaches a saturation value for
larger heating of the particle, indicating that $V_p$ does not
increase sufficiently faster as compared to the decrease of
$\tau_r$. On the other hand, if the propulsion velocity of the Janus
sphere exhibits a non-linear dependence (for the case when $c_{gs}=2,
c_{ps}=1$ and vice versa) on $\Delta T/T_0$,
we do not observe any saturation in  $\tau_r
V^2_p$.

In order to have a better insight into \cref{fig:effective_diff}~(a),
in particular, whether $\chi$ monotonically increases with $\Delta T/T_0$,
we look at the functional dependence of  $V_p,\tau_r$ and $\DHBM$ 
on $\Delta T/T_0$:
\begin{equation}
\begin{split}
& V_p=\tilde{\mu} (\Delta T/T_0) f(\Delta T/T_0), \quad \DHBM=D_0 g(\Delta T/T_0) \\
\quad &\textrm{and} \quad
 D_r=D^{r}_0 h(\Delta T/T_0)	.
\end{split}
\label{eq:V_p_Dr_D0}
\end{equation}
$D_0$ and $D^{r}_0$ are the translational and rotational diffusion constant 
of the colloid in the isothermal solvent. The functional forms of $g(x)$ and 
$h(x)$ follow from the theory of hot Brownian motion, whereas $f(x)$ follows 
from the measured propulsion velocity of the heated Janus colloid:
\begin{equation}
\begin{split}
& f(x)=(1+c_1 x), \quad  g(x) \approx 1+c_2 x \\
\quad &\textrm{and} \quad
 h(x) \approx 1+c_4 x .
\end{split}
\end{equation}
The coefficients $c_2,c_4$ can be exactly determined from the 
hydrodynamic theory of hot Brownian motion and
depend on the microscopic interactions of the solvent particles. 
\cite{Rings:2011gj,Rings2012a,Chakraborty:2011kk}
On the other hand  the coefficient $c_1$ which quantifies 
the nonlinear term in the propulsion velocity
depends on the microscopic interaction parameters 
between the solvent and the colloid(see \cref{fig:propulsion_vel}).
Rewriting the term $\chi \equiv V_p^2 \tau_r/3D\DHBM$ as
\begin{equation}
\chi\equiv V_p^2 \tau_r /3\DHBM= \frac{\tilde{\mu}^2 x^2f^2(x)}{6 D_0 D^{r}_0 g(x) h(x)}, 
\label{eq:rescaled_enhancement}
\end{equation}
with $x=\Delta T/T_0$. From \cref{eq:rescaled_enhancement} it becomes evident 
that a plot of $V_p^2 \tau_r/3 \tilde{\mu}^2\DHBM$ would exhibit a
data collapse for different choices $c_{\alpha \beta}$ 
whenever the non-linear increase in the propulsion velocity 
is not dominant. This happens when the heating of the colloid is 
small and in this regime 
$\chi \sim (\Delta T/T_0)^2$
This is indeed observed in the simulations (see \cref{fig:effective_diff}~(b)). 
For all the choices of the wetting parameters that we have
investigated, there is an initial data collapse and 
the initial increase in the $\chi$
is quadratic in  $\Delta T/T_0$. However, for the choice of 
$c_{gs}=2,c_{ps}=0$ and $c_{gs}=0,c_{ps}=2$, the propulsion velocity
increases linearly throughout the range of heating that we investigated 
($c_1=0$)
and consequently with increased heating $\chi \sim \Delta T/T_0 $ and 
approaches a constant value when the linear terms in $g(x)$ and $h(x)$
becomes relevant( the 
dashed line in \cref{fig:effective_diff}~(b). On the other hand, 
for the choice of $c_{gs}=2,c_{ps}=0$ and $c_{gs}=0,c_{ps}=2$, 
after the initial quadratic increase, there is a small regime when
the enhancement factor $\chi$ shows a cross over to a linear increase
in $\Delta T/T_0$ but eventually increases again as $(\Delta T/T_0)^2$ 
when the linear term in $f(x)$ starts to dominate. Although not investigated
in the current work, we note that from \cref{eq:rescaled_enhancement}, 
$\tilde{\mu} \sim R^{-1}$, $D_0 D^{r}_0 \sim R^4$ and $\chi \sim R^2$.

\section{Conclusion}
\label{sec:conclusion}
In conclusion, we have presented a simple microscopic model system
that can be used to investigate self-thermophoresis of a Janus
particle. The microscopic interfacial resistance across the
solid-liquid boundary is exploited to produce a self-created gradient
across the poles of the sphere. It is this self-created temperature
gradient which leads to a self-propulsion along the direction of the
symmetry axis. 
The rotational diffusion constant is extracted from the time
correlation of the symmetry axis
$\langle \bar{\vec{z}}(t)\bar{\vec{z}}(0) \rangle$ for different
heating's of the colloid, as well as for different choices of the
wetting parameter. We quantify the measured rotational diffusion $D_r$
against the temperature difference $\delta T$ across the poles of the
Janus sphere as well as the average surface temperature difference
$\Delta T$ from the ambient fluid. Since the rotational diffusion is
determined by the complete flow field in the solvent, we illustrate
that comparing $D_r$ against $\delta T$ is misleading and is better
quantified when compared against $\Delta T$ that results in a data
collapse for different choices of the microscopic interaction. The
average propulsion velocity is also measured for different choices of
the wetting parameter and the directionality of motion changes
depending on the microscopic interaction. We observe two important
feature in the average propulsion velocity -- first, for the choice of
repulsive interaction the phoretic mobility changes sign and
therefore, the direction of propulsion also reverses. This is also corroborated
by the probability distribution of the angle between the displacement
vector $\Delta \vec{r}(t) \equiv \vec{r}(t)-\vec{r}(0)$ and the
symmetry axis $\bar{\vec{z}}(0)$. And second, when
either half of the sphere has a repulsive interaction, the propulsion
velocity remains in the linear response regime even for higher heating
of the colloid. On the other hand, whenever the interaction of the
solvent with the colloid has an attractive term, we observe a
non-linear dependence of $V_p$ with the temperature difference
$\delta T$. This indicates that repulsive interaction suppresses the
phoretic mobility of the particle, and can be exploited in designing
artificial microswimmers. Further, the propulsion velocity exhibits a
threshold in heating below which there is no appreciable self-propulsion.
Finally, we combine the measured
propulsion velocity and the rotational diffusion time $\tau_r=1/2D_r$
to estimate the enhancement in the long time diffusion coefficient of
the particle $D_{\rm eff}=\DHBM\left(1+ V_p^2 \tau_r/3\DHBM\right)$.

\section{Acknowledgement} 
D.C acknowledges Klaus Kroy of University of Leipzig and 
Debasish Chaudhuri of IOP Bhubaneswar  for stimulating discussions and 
for a careful reading of the manuscript. This
work was supported by funding from the Science and Engineering
Research Board (SERB), India vide grant no. SB/S2/CMP-113/2013 and by
nVidia\textsuperscript{\textregistered} corporation through its GPU
Grant Program.

\section{References}
\bibliography{janus_refs}

\begin{thebibliography}{21}%
\makeatletter
\providecommand \@ifxundefined [1]{%
 \@ifx{#1\undefined}
}%
\providecommand \@ifnum [1]{%
 \ifnum #1\expandafter \@firstoftwo
 \else \expandafter \@secondoftwo
 \fi
}%
\providecommand \@ifx [1]{%
 \ifx #1\expandafter \@firstoftwo
 \else \expandafter \@secondoftwo
 \fi
}%
\providecommand \natexlab [1]{#1}%
\providecommand \enquote  [1]{``#1''}%
\providecommand \bibnamefont  [1]{#1}%
\providecommand \bibfnamefont [1]{#1}%
\providecommand \citenamefont [1]{#1}%
\providecommand \href@noop [0]{\@secondoftwo}%
\providecommand \href [0]{\begingroup \@sanitize@url \@href}%
\providecommand \@href[1]{\@@startlink{#1}\@@href}%
\providecommand \@@href[1]{\endgroup#1\@@endlink}%
\providecommand \@sanitize@url [0]{\catcode `\\12\catcode `\$12\catcode
  `\&12\catcode `\#12\catcode `\^12\catcode `\_12\catcode `\%12\relax}%
\providecommand \@@startlink[1]{}%
\providecommand \@@endlink[0]{}%
\providecommand \url  [0]{\begingroup\@sanitize@url \@url }%
\providecommand \@url [1]{\endgroup\@href {#1}{\urlprefix }}%
\providecommand \urlprefix  [0]{URL }%
\providecommand \Eprint [0]{\href }%
\providecommand \doibase [0]{http://dx.doi.org/}%
\providecommand \selectlanguage [0]{\@gobble}%
\providecommand \bibinfo  [0]{\@secondoftwo}%
\providecommand \bibfield  [0]{\@secondoftwo}%
\providecommand \translation [1]{[#1]}%
\providecommand \BibitemOpen [0]{}%
\providecommand \bibitemStop [0]{}%
\providecommand \bibitemNoStop [0]{.\EOS\space}%
\providecommand \EOS [0]{\spacefactor3000\relax}%
\providecommand \BibitemShut  [1]{\csname bibitem#1\endcsname}%
\let\auto@bib@innerbib\@empty
\bibitem [{\citenamefont {Popescu}\ \emph {et~al.}(2010)\citenamefont
  {Popescu}, \citenamefont {Dietrich}, \citenamefont {Tasinkevych},\ and\
  \citenamefont {Ralston}}]{Popescu:2010cq}%
  \BibitemOpen
  \bibfield  {author} {\bibinfo {author} {\bibfnamefont {M.~N.}\ \bibnamefont
  {Popescu}}, \bibinfo {author} {\bibfnamefont {S.}~\bibnamefont {Dietrich}},
  \bibinfo {author} {\bibfnamefont {M.}~\bibnamefont {Tasinkevych}}, \ and\
  \bibinfo {author} {\bibfnamefont {J.}~\bibnamefont {Ralston}},\ }\href
  {\doibase 10.1140/epje/i2010-10593-3} {\bibfield  {journal} {\bibinfo
  {journal} {The European Physical Journal E}\ }\textbf {\bibinfo {volume}
  {31}},\ \bibinfo {pages} {351--367} (\bibinfo {year} {2010})}\BibitemShut
  {NoStop}%
\bibitem [{\citenamefont {Popescu}, \citenamefont {Tasinkevych},\ and\
  \citenamefont {Dietrich}(2011)}]{Popescu:2011kf}%
  \BibitemOpen
  \bibfield  {author} {\bibinfo {author} {\bibfnamefont {M.~N.}\ \bibnamefont
  {Popescu}}, \bibinfo {author} {\bibfnamefont {M.}~\bibnamefont
  {Tasinkevych}}, \ and\ \bibinfo {author} {\bibfnamefont {S.}~\bibnamefont
  {Dietrich}},\ }\href {\doibase 10.1209/0295-5075/95/28004} {\bibfield
  {journal} {\bibinfo  {journal} {EPL (Europhysics Letters)}\ }\textbf
  {\bibinfo {volume} {95}},\ \bibinfo {pages} {28004} (\bibinfo {year}
  {2011})}\BibitemShut {NoStop}%
\bibitem [{\citenamefont {Golestanian}, \citenamefont {Liverpool},\ and\
  \citenamefont {Ajdari}(2007)}]{Golestanian:2007hu}%
  \BibitemOpen
  \bibfield  {author} {\bibinfo {author} {\bibfnamefont {R.}~\bibnamefont
  {Golestanian}}, \bibinfo {author} {\bibfnamefont {T.~B.}\ \bibnamefont
  {Liverpool}}, \ and\ \bibinfo {author} {\bibfnamefont {A.}~\bibnamefont
  {Ajdari}},\ }\href {\doibase 10.1088/1367-2630/9/5/126} {\bibfield  {journal}
  {\bibinfo  {journal} {New Journal of Physics}\ }\textbf {\bibinfo {volume}
  {9}},\ \bibinfo {pages} {126--126} (\bibinfo {year} {2007})}\BibitemShut
  {NoStop}%
\bibitem [{\citenamefont {Jiang}, \citenamefont {Yoshinaga},\ and\
  \citenamefont {Sano}(2010)}]{Jiang:2010el}%
  \BibitemOpen
  \bibfield  {author} {\bibinfo {author} {\bibfnamefont {H.-R.}\ \bibnamefont
  {Jiang}}, \bibinfo {author} {\bibfnamefont {N.}~\bibnamefont {Yoshinaga}}, \
  and\ \bibinfo {author} {\bibfnamefont {M.}~\bibnamefont {Sano}},\ }\href
  {\doibase 10.1103/PhysRevLett.105.268302} {\bibfield  {journal} {\bibinfo
  {journal} {Physical Review Letters}\ }\textbf {\bibinfo {volume} {105}}
  (\bibinfo {year} {2010}),\ 10.1103/PhysRevLett.105.268302}\BibitemShut
  {NoStop}%
\bibitem [{\citenamefont {Schoen}\ \emph {et~al.}(2006)\citenamefont {Schoen},
  \citenamefont {Walther}, \citenamefont {Arcidiacono}, \citenamefont
  {Poulikakos},\ and\ \citenamefont {Koumoutsakos}}]{Schoen:2006km}%
  \BibitemOpen
  \bibfield  {author} {\bibinfo {author} {\bibfnamefont {P.~A.~E.}\
  \bibnamefont {Schoen}}, \bibinfo {author} {\bibfnamefont {J.~H.}\
  \bibnamefont {Walther}}, \bibinfo {author} {\bibfnamefont {S.}~\bibnamefont
  {Arcidiacono}}, \bibinfo {author} {\bibfnamefont {D.}~\bibnamefont
  {Poulikakos}}, \ and\ \bibinfo {author} {\bibfnamefont {P.}~\bibnamefont
  {Koumoutsakos}},\ }\href {\doibase 10.1021/nl060982r} {\bibfield  {journal}
  {\bibinfo  {journal} {Nano Letters}\ }\textbf {\bibinfo {volume} {6}},\
  \bibinfo {pages} {1910--1917} (\bibinfo {year} {2006})}\BibitemShut {NoStop}%
\bibitem [{\citenamefont {Chen}(2000)}]{Chen:2000jv}%
  \BibitemOpen
  \bibfield  {author} {\bibinfo {author} {\bibfnamefont {S.}~\bibnamefont
  {Chen}},\ }\href {\doibase 10.1006/jcis.1999.6641} {\bibfield  {journal}
  {\bibinfo  {journal} {Journal of Colloid and Interface Science}\ }\textbf
  {\bibinfo {volume} {224}},\ \bibinfo {pages} {63--75} (\bibinfo {year}
  {2000})}\BibitemShut {NoStop}%
\bibitem [{\citenamefont {Howse}\ \emph {et~al.}(2007)\citenamefont {Howse},
  \citenamefont {Jones}, \citenamefont {Ryan}, \citenamefont {Gough},
  \citenamefont {Vafabakhsh},\ and\ \citenamefont
  {Golestanian}}]{Howse:2007ed}%
  \BibitemOpen
  \bibfield  {author} {\bibinfo {author} {\bibfnamefont {J.}~\bibnamefont
  {Howse}}, \bibinfo {author} {\bibfnamefont {R.}~\bibnamefont {Jones}},
  \bibinfo {author} {\bibfnamefont {A.}~\bibnamefont {Ryan}}, \bibinfo {author}
  {\bibfnamefont {T.}~\bibnamefont {Gough}}, \bibinfo {author} {\bibfnamefont
  {R.}~\bibnamefont {Vafabakhsh}}, \ and\ \bibinfo {author} {\bibfnamefont
  {R.}~\bibnamefont {Golestanian}},\ }\href {\doibase
  10.1103/PhysRevLett.99.048102} {\bibfield  {journal} {\bibinfo  {journal}
  {Physical Review Letters}\ }\textbf {\bibinfo {volume} {99}},\ \bibinfo
  {pages} {048102} (\bibinfo {year} {2007})}\BibitemShut {NoStop}%
\bibitem [{\citenamefont {Chakraborty}\ \emph {et~al.}(2011)\citenamefont
  {Chakraborty}, \citenamefont {Gnann}, \citenamefont {Rings}, \citenamefont
  {Glaser}, \citenamefont {Otto}, \citenamefont {Cichos},\ and\ \citenamefont
  {Kroy}}]{Chakraborty:2011kk}%
  \BibitemOpen
  \bibfield  {author} {\bibinfo {author} {\bibfnamefont {D.}~\bibnamefont
  {Chakraborty}}, \bibinfo {author} {\bibfnamefont {M.~V.}\ \bibnamefont
  {Gnann}}, \bibinfo {author} {\bibfnamefont {D.}~\bibnamefont {Rings}},
  \bibinfo {author} {\bibfnamefont {J.}~\bibnamefont {Glaser}}, \bibinfo
  {author} {\bibfnamefont {F.}~\bibnamefont {Otto}}, \bibinfo {author}
  {\bibfnamefont {F.}~\bibnamefont {Cichos}}, \ and\ \bibinfo {author}
  {\bibfnamefont {K.}~\bibnamefont {Kroy}},\ }\href {\doibase
  10.1209/0295-5075/96/60009} {\bibfield  {journal} {\bibinfo  {journal} {EPL
  (Europhysics Letters)}\ }\textbf {\bibinfo {volume} {96}},\ \bibinfo {pages}
  {60009} (\bibinfo {year} {2011})}\BibitemShut {NoStop}%
\bibitem [{\citenamefont {Rings}\ \emph {et~al.}(2011)\citenamefont {Rings},
  \citenamefont {Selmke}, \citenamefont {Cichos},\ and\ \citenamefont
  {Kroy}}]{Rings:2011gj}%
  \BibitemOpen
  \bibfield  {author} {\bibinfo {author} {\bibfnamefont {D.}~\bibnamefont
  {Rings}}, \bibinfo {author} {\bibfnamefont {M.}~\bibnamefont {Selmke}},
  \bibinfo {author} {\bibfnamefont {F.}~\bibnamefont {Cichos}}, \ and\ \bibinfo
  {author} {\bibfnamefont {K.}~\bibnamefont {Kroy}},\ }\href {\doibase
  10.1039/c0sm00854k} {\bibfield  {journal} {\bibinfo  {journal} {Soft Matter}\
  }\textbf {\bibinfo {volume} {7}},\ \bibinfo {pages} {3441} (\bibinfo {year}
  {2011})}\BibitemShut {NoStop}%
\bibitem [{\citenamefont {Merabia}\ \emph {et~al.}(2009)\citenamefont
  {Merabia}, \citenamefont {Keblinski}, \citenamefont {Joly}, \citenamefont
  {Lewis},\ and\ \citenamefont {Barrat}}]{Merabia:2009fk}%
  \BibitemOpen
  \bibfield  {author} {\bibinfo {author} {\bibfnamefont {S.}~\bibnamefont
  {Merabia}}, \bibinfo {author} {\bibfnamefont {P.}~\bibnamefont {Keblinski}},
  \bibinfo {author} {\bibfnamefont {L.}~\bibnamefont {Joly}}, \bibinfo {author}
  {\bibfnamefont {L.}~\bibnamefont {Lewis}}, \ and\ \bibinfo {author}
  {\bibfnamefont {J.-L.}\ \bibnamefont {Barrat}},\ }\href {\doibase
  10.1103/PhysRevE.79.021404} {\bibfield  {journal} {\bibinfo  {journal}
  {Physical Review E}\ }\textbf {\bibinfo {volume} {79}} (\bibinfo {year}
  {2009}),\ 10.1103/PhysRevE.79.021404}\BibitemShut {NoStop}%
\bibitem [{\citenamefont {Yang}, \citenamefont {Wysocki},\ and\ \citenamefont
  {Ripoll}(2014)}]{Yang2014}%
  \BibitemOpen
  \bibfield  {author} {\bibinfo {author} {\bibfnamefont {M.}~\bibnamefont
  {Yang}}, \bibinfo {author} {\bibfnamefont {A.}~\bibnamefont {Wysocki}}, \
  and\ \bibinfo {author} {\bibfnamefont {M.}~\bibnamefont {Ripoll}},\ }\href
  {\doibase 10.1039/C4SM00621F} {\bibfield  {journal} {\bibinfo  {journal}
  {Soft Matter}\ }\textbf {\bibinfo {volume} {10}},\ \bibinfo {pages} {6208}
  (\bibinfo {year} {2014})}\BibitemShut {NoStop}%
\bibitem [{\citenamefont {Yang}\ and\ \citenamefont
  {Ripoll}(2013)}]{Yang2013a}%
  \BibitemOpen
  \bibfield  {author} {\bibinfo {author} {\bibfnamefont {M.}~\bibnamefont
  {Yang}}\ and\ \bibinfo {author} {\bibfnamefont {M.}~\bibnamefont {Ripoll}},\
  }\href {\doibase 10.1039/c3sm27949a} {\bibfield  {journal} {\bibinfo
  {journal} {Soft Matter}\ }\textbf {\bibinfo {volume} {9}},\ \bibinfo {pages}
  {4661--4671} (\bibinfo {year} {2013})}\BibitemShut {NoStop}%
\bibitem [{\citenamefont {Yang}\ and\ \citenamefont
  {Ripoll}(2011)}]{Yang2011a}%
  \BibitemOpen
  \bibfield  {author} {\bibinfo {author} {\bibfnamefont {M.}~\bibnamefont
  {Yang}}\ and\ \bibinfo {author} {\bibfnamefont {M.}~\bibnamefont {Ripoll}},\
  }\href {\doibase 10.1103/PhysRevE.84.061401} {\bibfield  {journal} {\bibinfo
  {journal} {Physical Review E}\ }\textbf {\bibinfo {volume} {84}},\ \bibinfo
  {pages} {061401} (\bibinfo {year} {2011})}\BibitemShut {NoStop}%
\bibitem [{\citenamefont {Bickel}, \citenamefont {Zecua},\ and\ \citenamefont
  {W{\"{u}}rger}(2014)}]{Bickel2014a}%
  \BibitemOpen
  \bibfield  {author} {\bibinfo {author} {\bibfnamefont {T.}~\bibnamefont
  {Bickel}}, \bibinfo {author} {\bibfnamefont {G.}~\bibnamefont {Zecua}}, \
  and\ \bibinfo {author} {\bibfnamefont {A.}~\bibnamefont {W{\"{u}}rger}},\
  }\href {\doibase 10.1103/PhysRevE.89.050303} {\bibfield  {journal} {\bibinfo
  {journal} {Physical Review E - Statistical, Nonlinear, and Soft Matter
  Physics}\ }\textbf {\bibinfo {volume} {89}},\ \bibinfo {pages} {1--5}
  (\bibinfo {year} {2014})}\BibitemShut {NoStop}%
\bibitem [{\citenamefont {Vladkov}\ and\ \citenamefont
  {Barrat}(2006)}]{Vladkov:2006ema}%
  \BibitemOpen
  \bibfield  {author} {\bibinfo {author} {\bibfnamefont {M.}~\bibnamefont
  {Vladkov}}\ and\ \bibinfo {author} {\bibfnamefont {J.-L.}\ \bibnamefont
  {Barrat}},\ }\href {\doibase 10.1021/nl060670o} {\bibfield  {journal}
  {\bibinfo  {journal} {Nano Letters}\ }\textbf {\bibinfo {volume} {6}},\
  \bibinfo {pages} {1224--1228} (\bibinfo {year} {2006})}\BibitemShut {NoStop}%
\bibitem [{\citenamefont {Barrat}\ and\ \citenamefont
  {Chiaruttini}(2003)}]{Barrat:2003dx}%
  \BibitemOpen
  \bibfield  {author} {\bibinfo {author} {\bibfnamefont {J.-L.}\ \bibnamefont
  {Barrat}}\ and\ \bibinfo {author} {\bibfnamefont {F.}~\bibnamefont
  {Chiaruttini}},\ }\href {\doibase 10.1080/0026897031000068578} {\bibfield
  {journal} {\bibinfo  {journal} {Molecular Physics}\ }\textbf {\bibinfo
  {volume} {101}},\ \bibinfo {pages} {1605--1610} (\bibinfo {year}
  {2003})}\BibitemShut {NoStop}%
\bibitem [{\citenamefont {ten Hagen}, \citenamefont {van Teeffelen},\ and\
  \citenamefont {L{\"{o}}wen}(2011)}]{Teeffelen2010}%
  \BibitemOpen
  \bibfield  {author} {\bibinfo {author} {\bibfnamefont {B.}~\bibnamefont {ten
  Hagen}}, \bibinfo {author} {\bibfnamefont {S.}~\bibnamefont {van Teeffelen}},
  \ and\ \bibinfo {author} {\bibfnamefont {H.}~\bibnamefont {L{\"{o}}wen}},\
  }\href {\doibase 10.1088/0953-8984/23/19/194119} {\bibfield  {journal}
  {\bibinfo  {journal} {Journal of Physics: Condensed Matter}\ }\textbf
  {\bibinfo {volume} {23}},\ \bibinfo {pages} {194119} (\bibinfo {year}
  {2011})}\BibitemShut {NoStop}%
\bibitem [{\citenamefont {Rings}, \citenamefont {Chakraborty},\ and\
  \citenamefont {Kroy}(2012)}]{Rings2012a}%
  \BibitemOpen
  \bibfield  {author} {\bibinfo {author} {\bibfnamefont {D.}~\bibnamefont
  {Rings}}, \bibinfo {author} {\bibfnamefont {D.}~\bibnamefont {Chakraborty}},
  \ and\ \bibinfo {author} {\bibfnamefont {K.}~\bibnamefont {Kroy}},\ }\href
  {\doibase 10.1088/1367-2630/14/5/053012} {\bibfield  {journal} {\bibinfo
  {journal} {New Journal of Physics}\ }\textbf {\bibinfo {volume} {14}},\
  \bibinfo {pages} {053012} (\bibinfo {year} {2012})}\BibitemShut {NoStop}%
\bibitem [{\citenamefont {Auschra}\ \emph {et~al.}(2017)\citenamefont
  {Auschra}, \citenamefont {Falasco}, \citenamefont {Chakraborty},
  \citenamefont {Pfaller},\ and\ \citenamefont {Kroy}}]{Auschra2017}%
  \BibitemOpen
  \bibfield  {author} {\bibinfo {author} {\bibfnamefont {S.}~\bibnamefont
  {Auschra}}, \bibinfo {author} {\bibfnamefont {G.}~\bibnamefont {Falasco}},
  \bibinfo {author} {\bibfnamefont {D.}~\bibnamefont {Chakraborty}}, \bibinfo
  {author} {\bibfnamefont {R.}~\bibnamefont {Pfaller}}, \ and\ \bibinfo
  {author} {\bibfnamefont {K.}~\bibnamefont {Kroy}},\ }\href@noop {} {\enquote
  {\bibinfo {title} {Coarse graining nonisothermal active colloids},}\ }
  (\bibinfo {year} {2017}),\ \bibinfo {note} {unpublished Work}\BibitemShut
  {NoStop}%
\bibitem [{\citenamefont {Brilliantov}, \citenamefont {Denisov},\ and\
  \citenamefont {Krapivsky}(1991)}]{Brilliantov1991}%
  \BibitemOpen
  \bibfield  {author} {\bibinfo {author} {\bibfnamefont {N.~V.}\ \bibnamefont
  {Brilliantov}}, \bibinfo {author} {\bibfnamefont {V.~P.}\ \bibnamefont
  {Denisov}}, \ and\ \bibinfo {author} {\bibfnamefont {P.~L.}\ \bibnamefont
  {Krapivsky}},\ }\href {\doibase 10.1016/0378-4371(91)90408-5} {\bibfield
  {journal} {\bibinfo  {journal} {Physica A: Statistical Mechanics and its
  Applications}\ }\textbf {\bibinfo {volume} {175}},\ \bibinfo {pages}
  {293--304} (\bibinfo {year} {1991})}\BibitemShut {NoStop}%
\bibitem [{\citenamefont {L\"{u}sebrink}(2011)}]{Luesebrink2011}%
  \BibitemOpen
  \bibfield  {author} {\bibinfo {author} {\bibfnamefont {D.}~\bibnamefont
  {L\"{u}sebrink}},\ }\emph {\bibinfo {title} {Colloidal Suspensions in
  Temperature Gradients with Mesoscopic Simulations}},\ \href@noop {} {Ph.D.
  thesis},\ \bibinfo  {school} {Universit\"{a}t zu K\"{o}ln} (\bibinfo {year}
  {2011})\BibitemShut {NoStop}%
\end{thebibliography}%
\end{document}